\documentclass[pre,twocolumn,showpacs,aps]{revtex4-1}

\usepackage{amssymb,amsfonts,amsmath}
\usepackage[pdftex]{graphicx}
\usepackage{multirow}
\usepackage{color}

\begin{document}

\title{Systemic delay propagation in the US airport network}

\author{Pablo Fleurquin}\affiliation{Instituto de F\'{\i}sica Interdisciplinar y Sistemas Complejos IFISC (CSIC-UIB),\\ 07122 Palma de Mallorca, Spain,\\ Innaxis Foundation \& Research Institute, Jos\'e Ortega y Gasset 20, 28006 Madrid, Spain}
\author{Jos\'e J. Ramasco}\email{jramasco@ifisc.uib-csic.es}\affiliation{Instituto de F\'{\i}sica Interdisciplinar y Sistemas Complejos IFISC (CSIC-UIB),\\ 07122 Palma de Mallorca, Spain}
\author{Victor M. Eguiluz}\affiliation{Instituto de F\'{\i}sica Interdisciplinar y Sistemas Complejos IFISC (CSIC-UIB),\\ 07122 Palma de Mallorca, Spain}

\widetext

\begin{abstract} 
Technologically driven transport systems are characterized by a networked structure connecting operation centers and by a dynamics ruled by pre-established schedules. Schedules impose serious constraints on the timing of the operations, condition the allocation of resources and define a baseline to assess system performance. Here we study the performance of an air transportation system in terms of delays. Technical, operational or meteorological issues affecting some flights give rise to primary delays. When operations continue, such delays can propagate, magnify and eventually involve a significant part of the network. We define metrics able to quantify the level of network congestion and introduce a model that reproduces the delay propagation patterns observed in the U.S. performance data. Our results indicate that there is a non-negligible risk of systemic instability even under normal operating conditions. We also identify passenger and crew connectivity as the most relevant internal factor contributing to delay spreading.
\end{abstract}

\maketitle 

\section{Introduction}

Air transportation systems have been traditionally described as graphs with vertices representing airports and edges direct flights during a fixed time period~\cite{barrat04,guimera05}. These graphs are called airport networks and have been studied at different geographical resolution scales, restricted, for instance, to a single country (usually the U.S. (USAN)~\cite{opsahl08,gautreau09,wuellner10,lanci11} but also China~\cite{li04} or Europe~\cite{burghouwt05}), or for the whole world (WAN)~\cite{barrat04,guimera05}. These networks show high heterogeneity in the distribution of connections per airport and in the traffic sustained by each connection. A non linear relation between the number of connections of the airports (topology) and the number of passengers (traffic) has been observed in Ref.~\cite{barrat04} and used later for modelling~\cite{colizza07}. Furthermore, airport networks are structured in clusters of highly interconnected airports that reflect the geographical areas in which the traffic is naturally divided~\cite{sales07}. The dynamics of the connections and the traffic levels have been also analyzed for the USAN~\cite{gautreau09}. All of these are aspects of the graphs that influence their capability to transport persons, goods and even other less desirable passengers. For example the propagation of  infectious diseases at a global scale that occurs when infected persons travel across the network~\cite{hufnagel04,colizza06,balcan09,balcan09b,tizzoni12}. The modelling and forecasting of disease spreading patterns using air traffic data is a story of a notable success~\cite{balcan09,balcan09b,tizzoni12}. One can, thus, wonder if this success can be extended to the propagation of other phenomena. In particular, we are interested in considering here flight delays and the way in which congestion can become a systemic risk.

According to the $2008$ Report of the Congress Joint Economic Committee, flight delays have an economic impact in the U.S. equivalent to $40.7$ billions of dollars per year~\cite{congress}, while a similar cost is expected in Europe~\cite{iccsai,eurocontrol}. The situation can turn even grimmer in the next decade since the air traffic is envisaged to increase~\cite{congress,iccsai,eurocontrol,jetzki09}. Delays damage companies' balances due to enhanced operation costs contributing to deteriorate their image with costumers~\cite{folkes87}. Passengers suffer a loss of time, even more acute in case of missing connections, that translates into decreased productivity, missed business opportunities or leisure activities. Additionally, attempts to recover delays lead to excess fuel consumption and larger CO$_2$ emissions.  As a consequence of this challenging situation, a considerable effort has been invested in the area of Air Traffic Management to characterize the sources of initial (primary) delays~\cite{rupp07,ahmadbeygi08} and the way in which they may be transferred and amplified by consequent operations, the so-called reactionary delays~\cite{jetzki09,beatty99,mayer03,bonnefoy07,cook07,belobaba09}. The concept of delay itself implies a time difference with respect to the baseline provided by a predefined schedule~\cite{rupp07,mayer03}. The propagation of delays thus corresponds to the spreading of a malfunction across the system. The mechanisms responsible for it reflect the complexity of air traffic operations. Apart from the airport networks structure and dynamics, other factors contributing to the delay propagation are airport congestion~\cite{bonnefoy07},  plane rotation or crew and passenger connection disruptions~\cite{beatty99,cook07,belobaba09}. Airline schedules typically include a buffer time to deal with all these issues. However, when this time is not enough, the departure of the next flight gets delayed and can affect further operations in a cascade-like effect~\cite{beatty99}.
There has been several attempts to model delay spreading~\cite{schaefer01,rosenberg02,wang03,janic05,bazzan09,lacasa09}. These studies differ in the level of detail included but in general they consider the effects of delays or disruptions in the operations of a few major airports (hubs).
In this work, we take instead a network-wide perspective to analyze the performance of a transportation system.  We define metrics able to quantify the level of spread of the delays in the network. We then apply these metrics to a database with information on the operations in the U.S during $2010$,  and introduce a model that reproduces the delay propagation patterns observed in the data. The model shows also a notable capacity to evaluate the risk of development of system-wide congestion and to assess the resilience of daily schedules to service disruptions.

\begin{figure}
\begin{center}
\includegraphics[width=8.6cm]{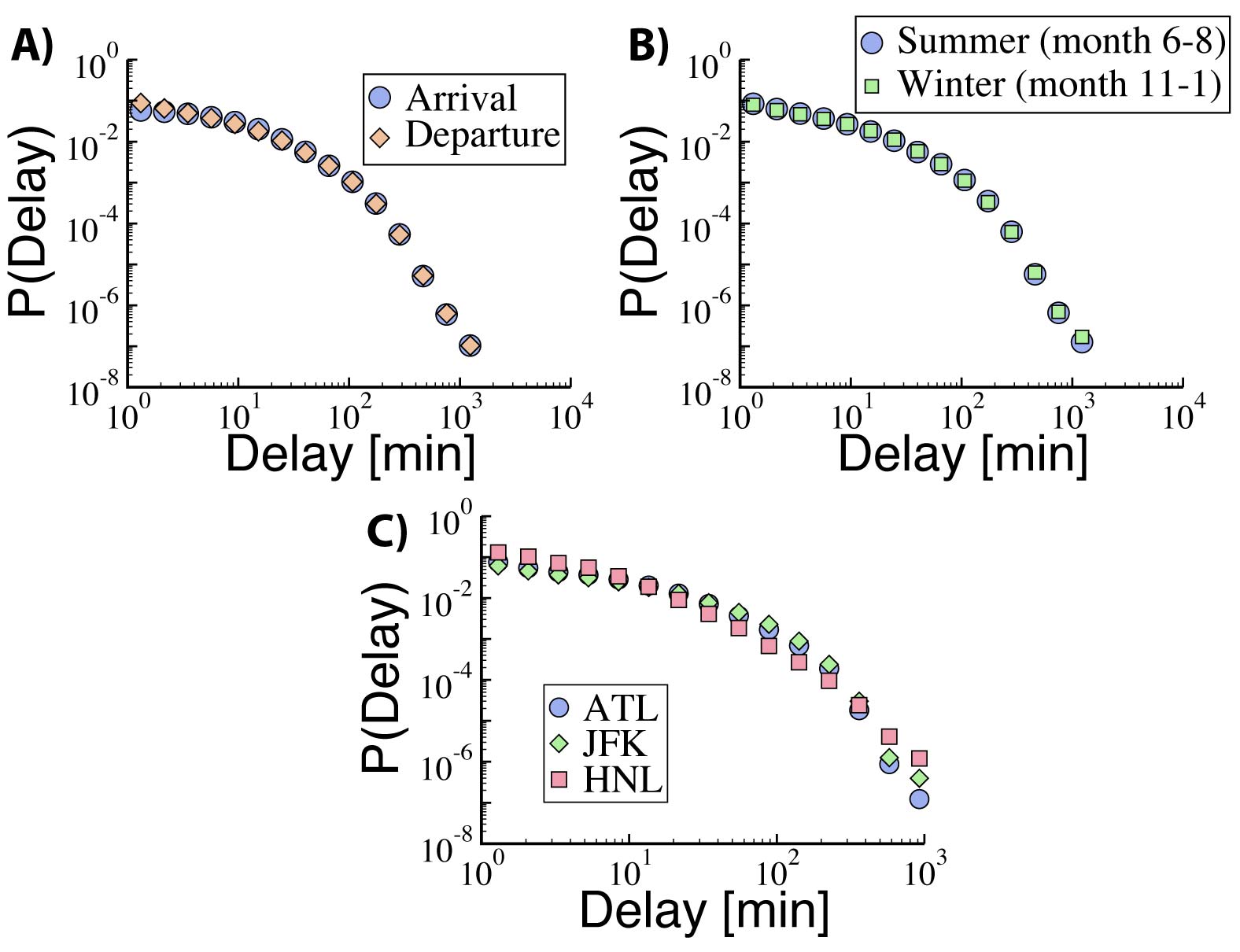}
\caption{Characterization of flight delays in the U.S. during 2010. A) Distribution of the delay per flight for arrivals and departures. B) Distribution of departure delays separating the flights according to the season: Summer and winter. C) Delay distribution for flights departing from Atlanta Hartsfield-Jackson (ATL), New York John F. Kennedy (JFK) and Honolulu (HNL) airports.}
\end{center}
\end{figure}

\section{Database}

The data was downloaded from the web page of the Bureau of Transport Statistics (BTS)~\cite{note2}. In particular, we used the Airline On-Time Performance Data, which is built with flight statistics provided by air carriers that exceed one percent of the annual national revenue for domestic regular service. The database comprehends $6,450,129$ scheduled flights operated by $18$ carriers connecting $305$ different commercial airports. The total flights operated in the US in $2010$, not only those that report on-time performance data, sum up $8,687,800$~\cite{bts-press}. Therefore, the database comprises information accounting for $74\%$ of them. The information per flight includes real and scheduled departure (arrival) times, origin and destination airport, an identification code (tail number) for each aircraft, airline, etc. This data enables us to represent the US airport network and furthermore replicate the scheduled flights for every day of $2010$. A detailed description can be found in Appendix A. It is important to note that this schedule is based on real events, which in some occasions may differ from the original planned schedule of the companies. If a flight gets canceled, diverted or even rescheduled the airline may introduce changes in the original schedule that are not possible to trace back. However, given that these flights represent, respectively, the $0.20\%$ and $1.75\%$ of all flights in the database, one can expect these changes not to be of large magnitude.

\section{Model}

The modeling approach followed is agent-based at the level of aircrafts and is data-driven in the sense that the daily schedules and the primary delays are obtained directly from real records in the database. This level of realism is necessary to confront the model predictions with the real unfolding of the delay events during each day. Concretely, the model dynamics simulates three main subprocesses: aircraft rotation, flight connectivity and airport congestion. The latter two are independent from each other, and can be turned on/off to explore the relevance of each subprocess in leading to network-wide congestion. Aircraft rotation, on the other hand, is intrinsic to the schedule and cannot be suppressed. 

\begin{figure}
\begin{center}
\includegraphics[width=8.6cm]{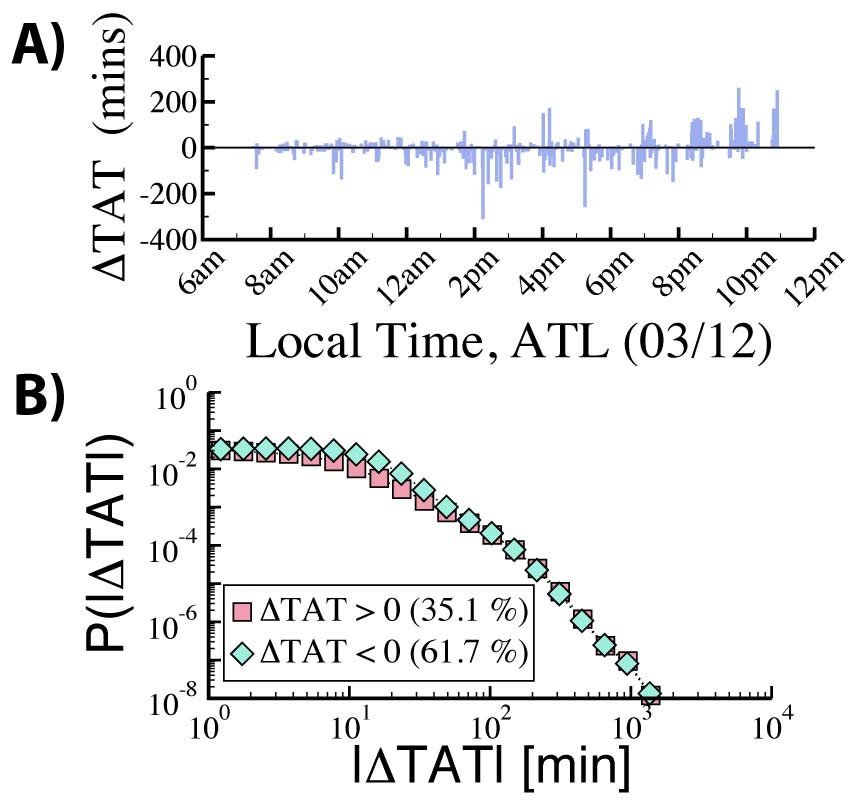}
\caption{In A), difference between the scheduled and real Turn Around Time ($\Delta TAT$) for operations in Atlanta airport, ATL, on March 12. In B), distribution of  the $\Delta TAT$ per flight, separating positive and negative contributions.}
\end{center}
\end{figure}

\begin{figure*}
\begin{center}
\includegraphics[width=17cm]{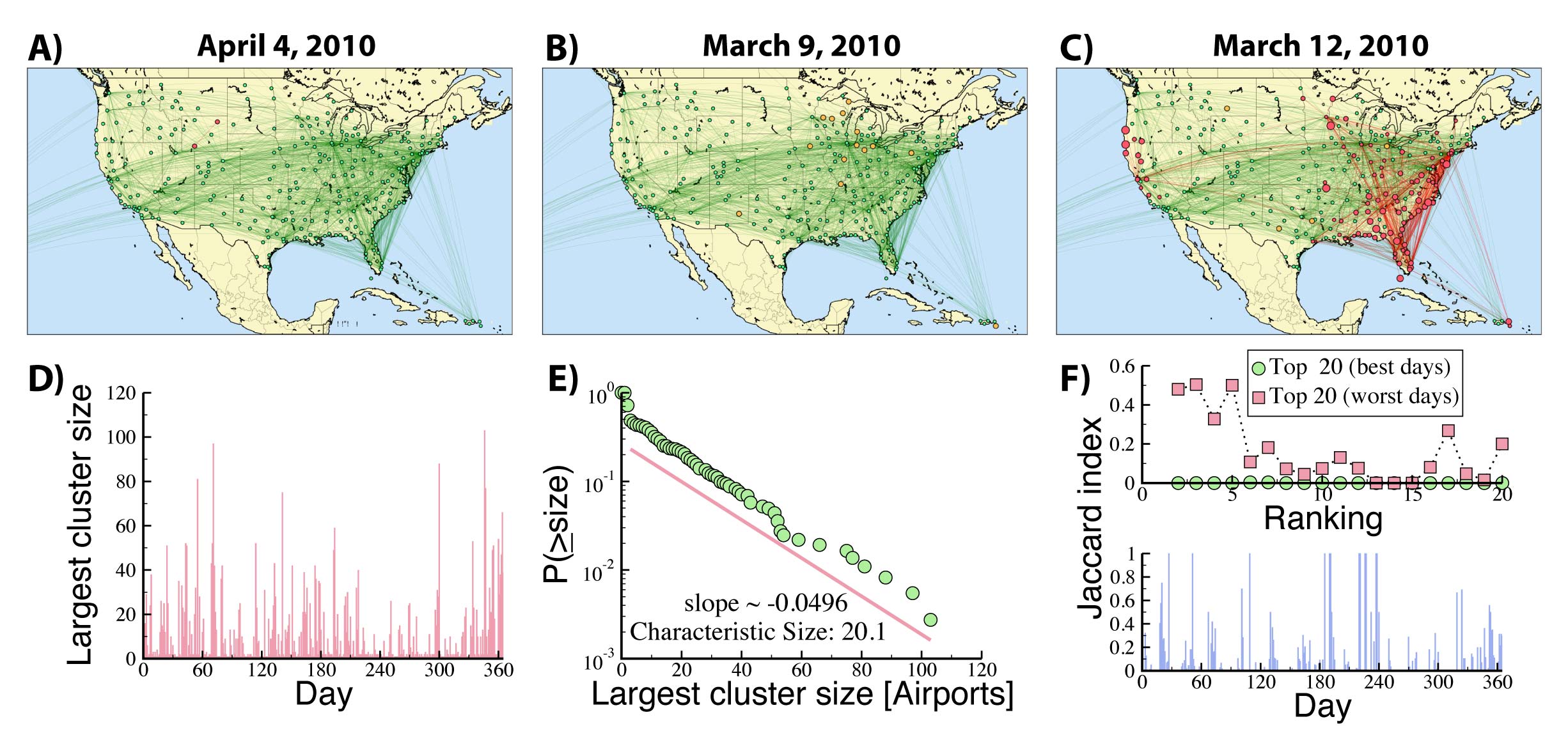}
\caption{Clusters of congested airports. Maps of the congested airports showing also connections between them for days with: A) low, B) intermediate, and C) high level of congestion. The airport color codes are: red, congested airport belonging to the largest cluster; orange, congested airport not belonging to the largest cluster; green, airport not congested. Links connecting airports in the largest cluster are in red. In D), daily size of the largest cluster as a function of time. In E), complementary cumulative distribution of the size of the largest cluster (log-normal scale). And in F), Jaccard index comparing airports belonging to the largest clusters in consecutive days or consecutive ranking positions according to the top 20 days with largest or lowest average delay.}
\end{center}
\end{figure*}

The basic time unit of the simulations is one minute, every aircraft state is tracked at this temporal resolution. We assume that the flights are not able to recover delays on air, and so the departure delays are equal to those at arrival to destination. Throughout a day, each aircraft follows the connections given in the schedule, the so-called plane rotations. The airports are supposed to have a capacity per hour proportional to the scheduled airport arrival rate with a proportionality factor $\beta$. Further arrivals produce delays. Passengers (crew) of incoming flights have a certain probability of connecting with other flights within a time window of $3$ hours from the scheduled arrival. The probability of connection is proportional, with a factor $\alpha$, to flight connectivity levels provided by the BTS for each U.S. airport. A more precise description of the model is included in Appendix B. This model has, thus, two free parameters: $\alpha$, controlling passenger connectivity, and $\beta$, accounting for airport capacity. In the following section, we will examine the effect of these parameters on the systemic spread of delays.

\begin{figure*}{b}
\begin{center}
\includegraphics[width=16cm]{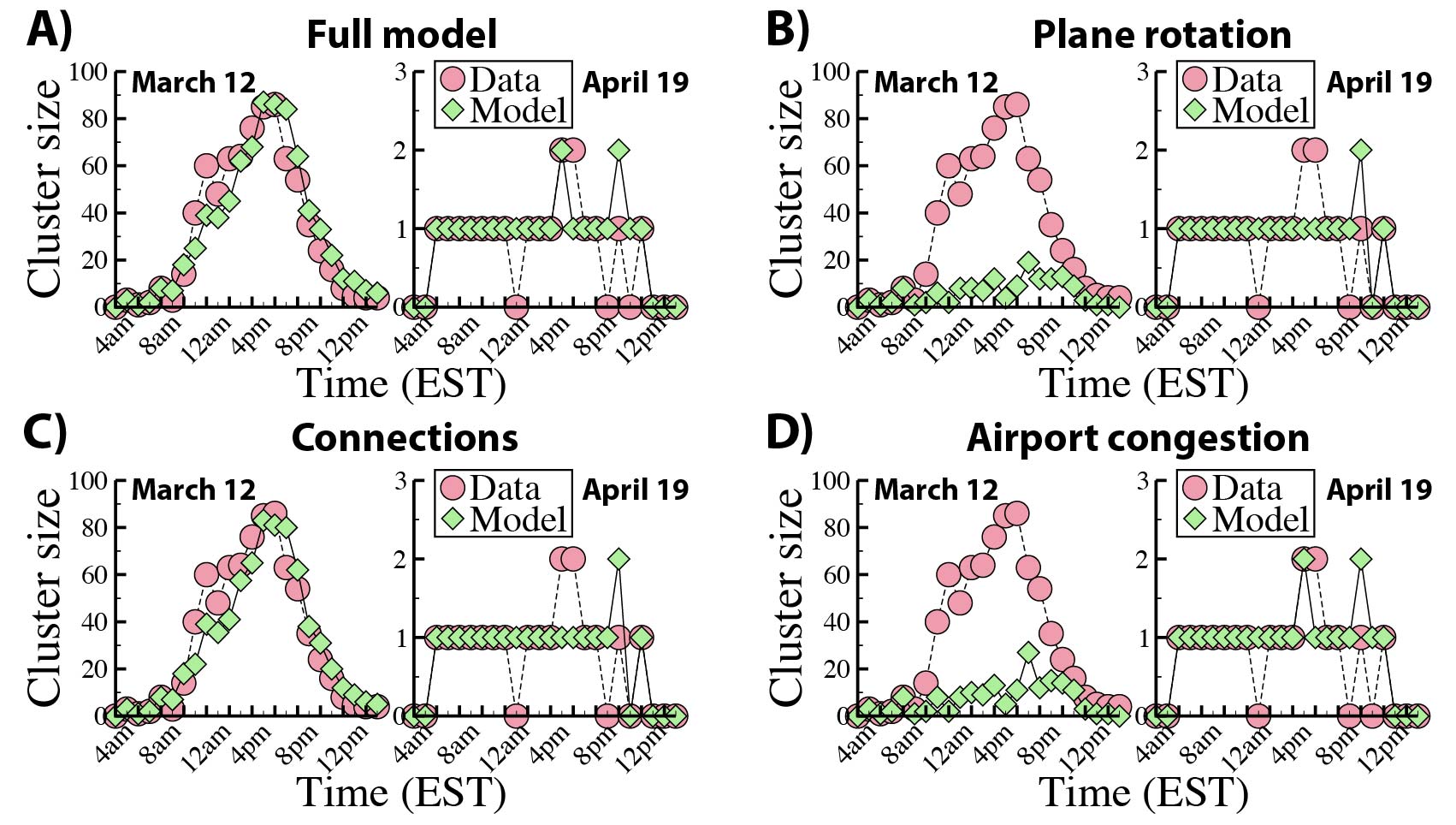}
\caption{Comparison model-reality. Evolution of the largest cluster per hour: A) for the full model, B) the model only with plane rotations, C) only with plane rotations and passenger connections and D) only with plane rotations and airport congestion. The selected days are the ones with the lowest delay (April $19$) and the second day with the largest delay (March $12$).}
\end{center}
\end{figure*}

\section{Data analysis and comparison with model predictions}  

Flight delays are defined as the difference between the scheduled and real departure (arrival) times~\cite{rupp07,mayer03}. Actually most of the flights operated in $2010$ were on time, even some before schedule, but $37.5 \%$ of those reporting performance arrived or departed late. Their delays do not show a characteristic value: the delay distribution displays a broad tail as can be seen in Figure 1A. This implies that most flights arrived late by just a few minutes, while others were hours behind schedule. The shape of the distributions is similar regardless of the arrival or departure nature of the operations. The planned buffer time on ground for each aircraft should help absorb part of the delays, specially those mildest as will be discussed next, thus altering the shape of the departure delay distribution. However, this factor is not able to substantially modify the characteristics of the distributions. Interestingly, the shape of the delay distribution does not change either when the season of the year is considered. Summer concentrates the major part of the year traffic, so the total delay is higher but when the distribution of delay per flight is taken into account both summer and winter behave similarly (see Figure 1B). The overall distributions of delays are thus quite robust. Some small differences can be only observed when focusing on particular airports. In Figure 1C, the departure delay distribution is plotted for Atlanta, JFK New York and Honolulu airports. While the distributions in Atlanta and New York are similar, the Honolulu airport shows a bias toward larger delays due to its isolation from the continent.

The effect of the buffer time in the airports for absorbing delays can be measured using the Turn Around Time ($TAT$). The $TAT$ stands for the time spent by an aircraft on ground from arrival to departure from the gate. This measure is associated with airport operational efficiency and is used to improve the planning of flight connectivity and aircraft rotational sequence stability~\cite{Wu00}. We refer as $\Delta TAT$ to the difference between scheduled and real times at the gate. On the one hand, a negative value of $\Delta TAT$  means that an aircraft stayed at the gate longer than expected and so fresh delay was introduced. On the other hand, a positive $\Delta TAT$ shows that the operation was quicker than scheduled and that part of the delay was recovered. In Figure 2A, we depict $\Delta TAT$ for each flight along a day in the most trafficked airport of the network: Hartsfield-Jackson in Atlata (ATL). That day, March $12$, happened to be one of the worst in the database in terms of average flight delay. The abundance of positive values of $\Delta TAT$ is a prove in favor of the capacity of the airport to recover delays. The distributions of $\Delta TAT$ for all the operations in $2010$ separated in positive an negative values are displayed in Figure 2B. These distributions, as those for the delays, show long tails, which is a marker of the complex nature of delay spreading mechanisms.

The focus so far has been on individual flight delays. We define now a metric of congestion for the full network. To do so, the average delay of all delayed flights during the year is taken as baseline and amounts to $29$ minutes. An airport is considered as congested whenever the average delay of all its departing flights over a certain period of time exceeds $29$ minutes. Additionally, a daily airport network is built using the flights of the day to assess whether congested airports are organized in connected clusters or not. Note that being in the same cluster is a measure of spatio-temporal correlation of congestion but not necessarily a sign of a cause-effect relation. We apply the same metric in the simulations in order to compare empirical and model results. Maps with the congested airports and the connections between them are shown for different days of the database in Figures 3A-3C. As can be seen, the scenario dramatically changes from day to day: in some days a large cluster surges covering $1/3$ of all airports, while in others only one or two airports cluster together. This is confirmed when the size of the largest connected cluster is depicted as a function of the day in Figure 3D. A strong variability is thus the main characteristic of the dynamics of the size of the largest congested cluster. The cumulative distribution of the cluster size is displayed in Figure 3E and it seems compatible with an exponential decay. Even if the fluctuations are large, there exists a well defined characteristic cluster size. Given the cluster variability, an important question to answer is whether the congested airports are recurrent. In panel 3F, we calculate the Jaccard index to compare the sets of airports in the largest cluster in consecutive days or for the top $20$ worst and best days. This index is $1$ if the clusters are equal and $0$ if they are strictly different. Interestingly, the index is relatively low for days with large clusters, which implies that the same airports are not consistently part of the cluster.

\begin{figure*}
\begin{center}
\includegraphics[width=16cm]{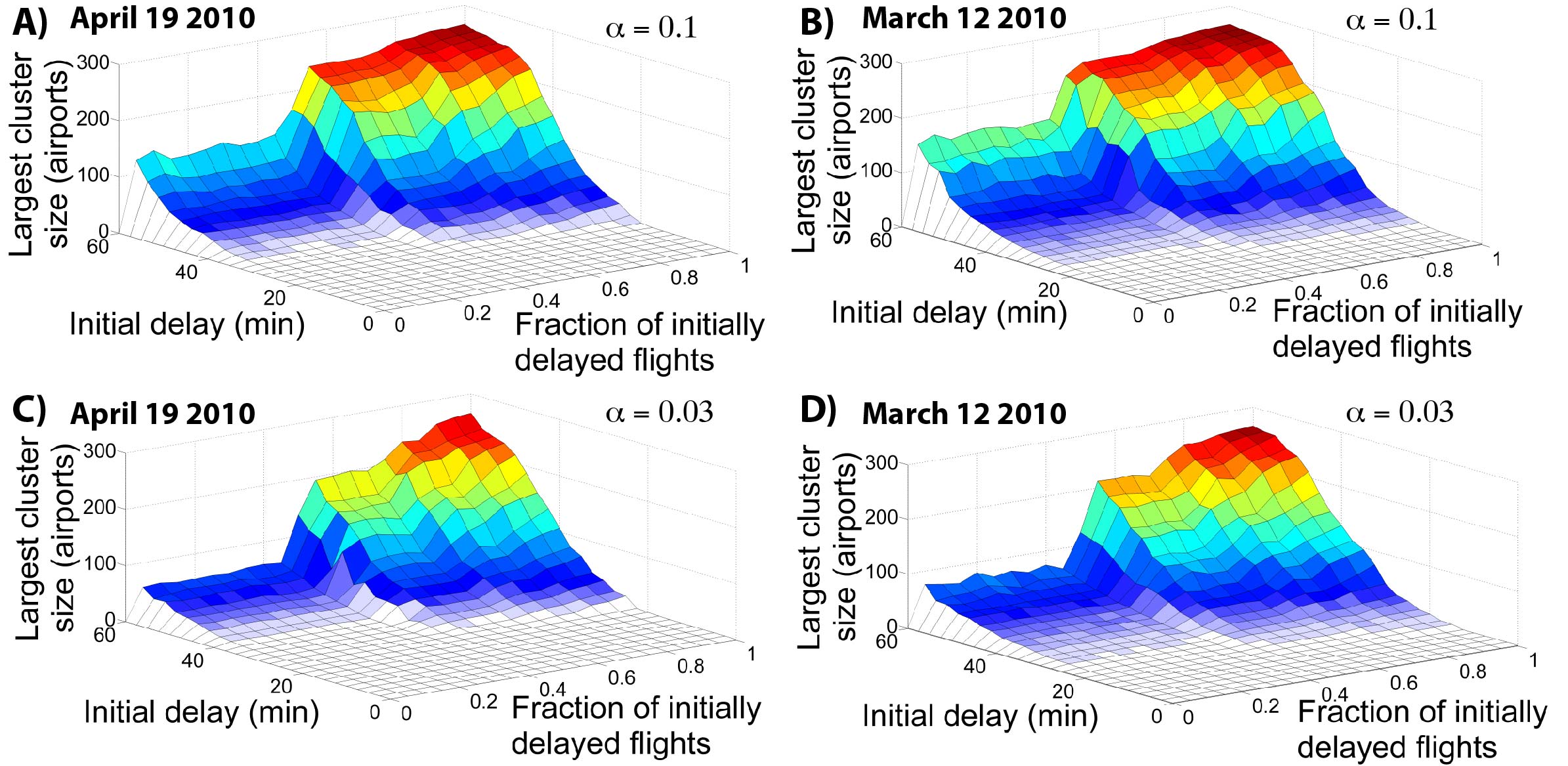}
\caption{Assessment of the schedule resilience to develop large clusters. In the plots, the size of the largest congested cluster is displayed as a function of the fraction of initial delayed flights and of the intensity of the initial delays for a congested March 12 in B) and D),  and for an uncongested day on April 19, A) and C), for two values of the flight connectivity factor $\alpha$. An initial fixed delay is assigned to randomly chosen flights.}
\end{center}
\end{figure*}

In order to compare empirical results and model predictions regarding the evolution of the cluster of congested airports, we run the model fixing the airport capacity parameter $\beta = 1$ and fitting the flight connectivity factor $\alpha$ to obtain a maximum cluster size similar to the one observed as the data. By fixing $\beta$ to $1$, we are assuming the same airport capacity as originally scheduled. The results for the temporal evolution of the congested cluster size hour by hour can be seen in Figure 4 for March $12$ and April $19$. Similar plots for other days of the year are included in Appendix C. Note that the fit of $\alpha$ is essential to get the maximum of these curves, however all the cluster size evolution predicted by the model follows strikingly well that of the real data. Actually, almost $60 \%$ of the airports in the real cluster are correctly identified by the model since they are top ranking when airports are ordered by probability of congestion. Furthermore, by fixing $\alpha$, without any fitting, the model can predict with $66 \%$ accuracy if a day will develop or not a large congested cluster (see Appendix C for further details). The model allows us also to explore which are the contributions of the main three ingredients (plane rotation, flight connectivity or airport congestion) to propagate delays. From Figures 4B-C, we can conclude that flight connectivity is the most important factor. One may still wonder if the picture changes when the capacity of the airports is modified. Actually, the model exhibits weak sensitivity to variations on the $\beta$ coefficient as shown in Figure 19. Slightly increasing the airport capacity will not ease off the propagation of delays since the main cause of the spreading, flight connections, is independent of it.  Conversely, a very strong decrease on the airports' capacity, around $50 \%$, is needed to trigger new primary delays that later on will spread in a cascading effect. This might be the case when generalized severe weather conditions or labor conflicts occur.  

The initial delays affect the outcome of the model. In the results of Figure 4, we take the primary delays for each aircraft from the data as initial conditions for the model. Introducing different initial conditions, we can assess the resilience of a day schedule to an increase of unexpected incidences. This question is explored in Figure 5 where a fraction of randomly selected flights are delayed. The size of the largest cluster is estimated as a function of the fraction of delayed flights and of the intensity of the initial delays. For the sake of simplicity, we set all the initial delays in the simulation equal to a fixed value (delay intensity in Figure 5). The results are displayed for the schedules of two days: April $19$ and March $12$, which respectively show a very small and very large cluster in the real data. In particular, the average flight delay on March $12$ was the second largest in $2010$. The congestion on the worst day of the year, October $27$, can be explained due to extreme meteorological conditions~\cite{note}, while on March $12$ no major external event was reported. Therefore, the network-wide propagation of delays in that day was likely caused and driven by internal mechanisms of the system. Comparing in Figure 5 the curves for March $12$ and April $9$, one notices that the surface representing the largest cluster size for March $12$ are displaced toward smaller values of the initial delay intensity or fraction of flights with primary delay. This shows a higher susceptibility of the schedule of this day to disruptive perturbations. Another interesting feature of the curves of Figure 5 is that, given enough primary delays, they show a non-negligible risk of systemic failure regardless of the schedule. The curves in Figure 5 for different values of $\alpha$ also confirm the relevance of connections and crew rotations for the spreading of delays.

\begin{figure}
\begin{center}
\includegraphics[width=8.6cm]{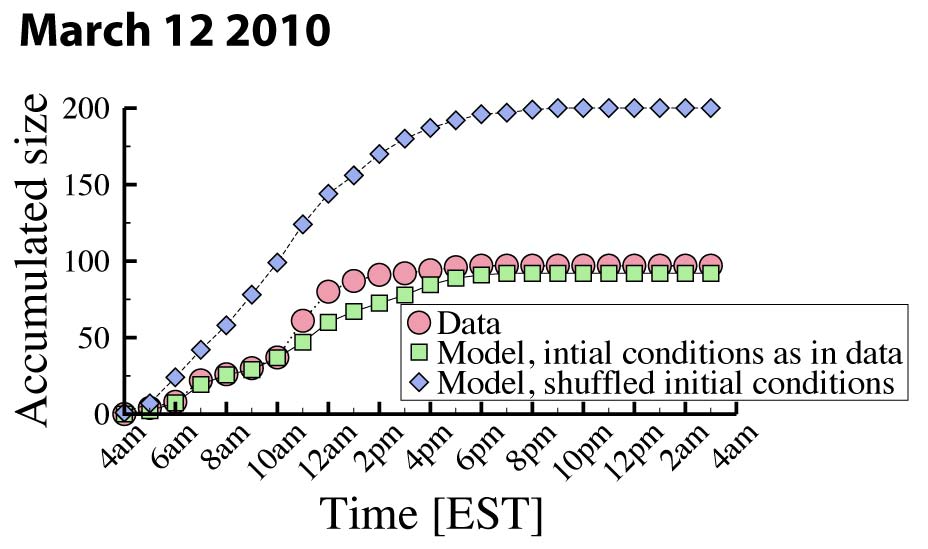}
\caption{Time evolution of the cumulative size of the largest congested cluster for different initial delays of the flights: assigned as found in data or to randomly selected flights but keeping the same values as in the data.}
\end{center}
\end{figure}

The primary flight delays in a day of real operations do not necessarily localize randomly in the network. If the causes are bad weather, technical or labor issues are more prone to concentrate in a few airports. In Figure 6, this issue is explored by comparing the intra-day evolution of the cumulative size of the largest congested cluster when the initial delays are introduced in the model in two different ways. The first one is by using the primary delays given in the database. The second procedure is by randomly shuffling the flights affected by the primary delays. The values of the real delays in the database are maintained but they are assigned to flights selected at random. The comparison of the curves for the two cases with the real data shows that random perturbations are way more efficient to collapse the system. While airports in general have some capacity to recover delays, the random selection of delayed flights affect a larger number of them and besides concentrate a heavier burden on smaller airports which have less capacity to react. This result evinces that the method followed for schedule evaluation in Figure 5 is conservative in the sense that it considers the schedule under a non favorable scenario for the distribution of primary delays.

\section{Discussion}  

In summary, we analyze the spreading of delays in an air traffic network. In particular, our results focus on the US airport network in $2010$ but the concepts and techniques employed can be easily extrapolated to the analysis of the performance of a generic transport system. We introduce a measure for the level of network-wide extension of the delays by defining when an airport is considered as congested and studying how congested airports form connected clusters in the network. The size of the largest congested cluster displays in the data a high variability from one day to the next. This feature is due to the re-start that the system suffers at the end of each day and points toward the relevance of the daily schedule to define the delay propagation patterns. In addition we introduce a data-driven model able to reproduce the delay evolution observed in the data. The model includes three main mechanisms to spread delays: Plane rotation, flight connections of either passenger or crews and airport congestion. The last two processes can be modulated at will to understand the role that each one of them plays in delay propagation. Our simulations evidence that passenger and crew connections is the most effective single mechanism to induce network congestion. We show how the model can be used to assess the daily schedule ability to deal with an increase in the number of disruptive events and also study the relevance of primary delay localization for the evolution of congestion in the network. Furthermore the model offers the possibility of evaluating the effects of interventions in the system before their real implementation.

Flight delays represent failures to meet constraints imposed by a daily schedule. Its propagation in the network is a paradigmatic example of the way in which a distributed transport system moves toward collapse. The framework develop in this work is thus of easy extension to system with dynamics regulated by predefined schedules. Its translation to other airport networks is, of course, straightforward, and even though the modeling of other transportation systems may require some particular details, the applicability of the metrics defined to measure network-wide congestion based on clustering is universal.

\begin{acknowledgments}
PF receives support from the network Complex World within the WPE of SESAR (Eurocontrol and EU Commission). JJR acknowledges funding from the Ram\'on y Cajal program of the Spanish Ministry of Economy (MINECO). Partial support from MINECO was also received through project MODASS (FIS2011-24785) and from the EU Commission through projects EUNOIA (FP7-DG.Connect-318367) and LASAGNE (FP7-ICT-318132).
\end{acknowledgments}

\appendix

\section{Database}

\subsection{Description}
\label{sec:DataNetworkFlSch}
The database was obtained using the information available at the Bureau of Transport Statistics~\cite{note2}. In particular, we used the Airline On-Time Performance Data, which is built with flight data provided by air carriers that exceed one percent of the annual national revenue for domestic scheduled service. The database comprehends $6,450,129$ scheduled domestic flights operated by $18$ carriers connecting $305$ different commercial airports. Considering all flights in $2010$, not only those that report On-Time Performance Data, the number of scheduled domestic flights totalizes $8,687,800$~\cite{bts-press} and so our dataset includes information for $74\%$ of the total. It is worth noting that this schedule is based on real events, which is not necessarily the original schedule in hands of the companies at the beginning of the day. If a flight gets canceled, diverted or even in milder situations the managers of an airline can introduce changes in the original schedule that we cannot trace back. However, given that these flights represent, respectively, the $0.20\%$ and $1.75\%$ of all flights in the database, one can expect these changes not to be of large magnitude. 

Among the available 
data fields, we consider the following as the most relevant for our work: \emph{Tail number, airline ID, airports of origin and destination, date of the flight, scheduled departure and arrival times, real departure and arrival times, and whether the flight was canceled or diverted}. The tail number is a code that identifies the aircraft and that allows us to follow it along the daily plane rotation. Arrival and departure times (real or scheduled) considered refer to when the flight actually reaches or departs from the gate, we are not taking taxing or take-off and landing times as departure (arrival) times. We will exclude of the coming analysis diverted and canceled flights since they difficult the characterization of the delay propagation and are a small fraction of the total operations.

\begin{figure}
\begin{center}
\includegraphics[width=8.6cm]{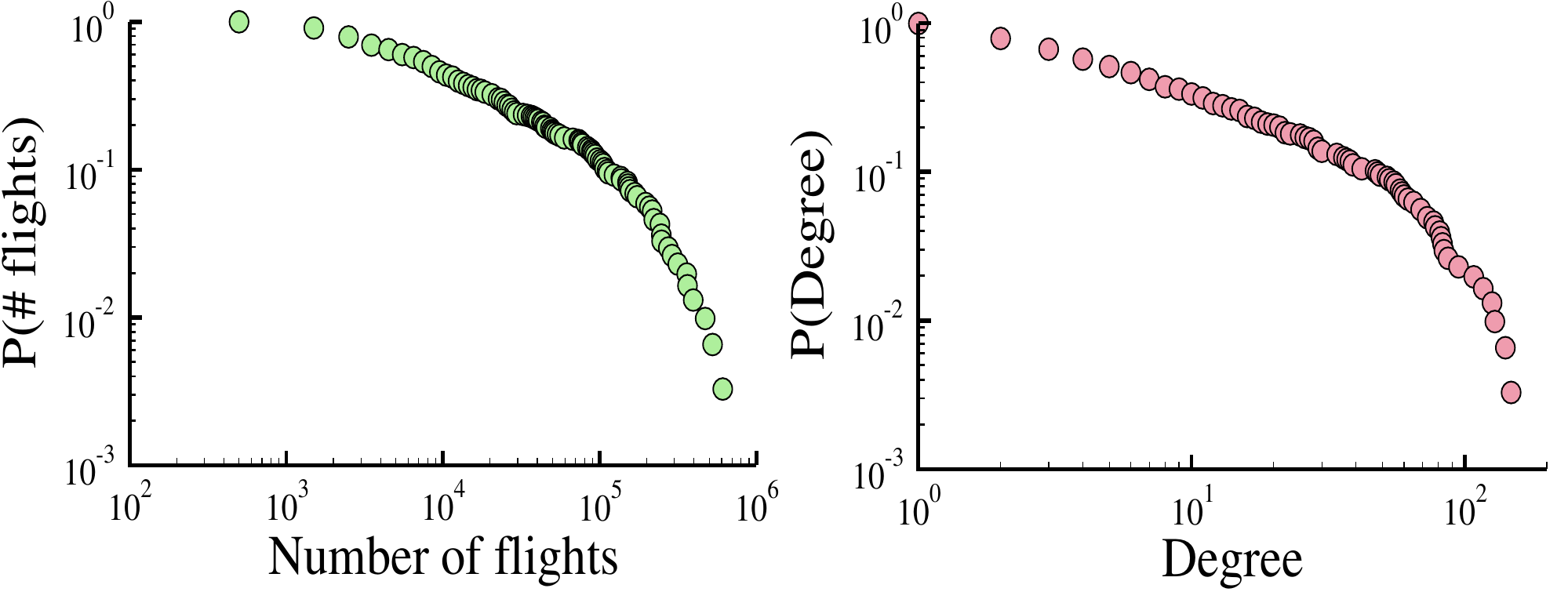}
\caption{Cumulative distribution of the number of flights and number of different connections (degree) for the airports in $2010$.}
\label{fig:Dist}
\end{center}
\end{figure}

A network between airports can be built based on the data: Airports are the vertices and edges represent direct flights from one airport to another. Note that this is a directed graph and that depends on the time-scale of data aggregation. On an annual basis, the resulting US air-transportation network comprises $305$  commercial airports and $2,318$ connections. Figure ~\ref{fig:Dist} depicts the complementary cumulative distribution of the number of flights and different connections for all the airports of the network in 2010. Both distributions confirm the presence of high heterogeneities in the airport network. The most active airports for $2010$ are represented in Table~\ref{tab:AirportDegree}, showing that the maximal degree corresponds to Atlanta International Airport (ATL) with $159$ different connections. For the analysis of the clusters of congested airports, we will considered networks aggregated only during one day.

\begin{table}[h]
\begin{center}
\begin{tabular}{c|c|c}
Airport code & \# edges & \# flights \\
\hline
\hline
ATL & $159$ & $809,869$  \\
ORD & $147$ & $608,981$  \\
DFW & $140$ & $524,206$  \\
DTW & $128$ & $314,369$  \\
DEN & $125$ & $470,592$  \\
MSP & $116$ & $246,245$  \\
IAH & $107$ & $362,562$  \\
SLC & $94$ & $246,245$  \\
MEM & $86$ & $152,730$  \\
MCO & $83$ & $241,851$  \\
\hline
\hline
\end{tabular}
\end{center}
\caption{Major airports according to their degree (number of different destinations).}
\label{tab:AirportDegree}
\end{table}

The average delay per delayed flights (those with positive delays) for $2010$ is $29$ minutes. This same value is used to define when an airport is considered congested in the main text but it can also be used to define days of operational problems or not. Those are respectively days whose average delay per delayed flight is over or below $29$ minutes, respectively. Table~\ref{tab:DayRanking} shows the ranking of the 20 best and worst days of the year according to their average delay for flights with positive delay.

\begin{table}[h]
\begin{center}
\begin{tabular}{c|c||c|c}
\multicolumn{2}{c}{Problematic days} & \multicolumn{2}{c}{Satisfactory days} \\
\hline
 DATE & Average delay (mins.) & DATE & Average delay (mins.)\\
\hline
\hline
Oct, $27$ & $54.3$  & Apr, $19$ & $16.9$\\
Mar, $12$ & $53.0$  & Oct, $09$ & $17.2$\\
Dec, $12$ & $51.9$  & Nov, $11$ & $17.3$ \\
Jan, $24$ & $49.8$  & Apr, $14$ & $17.6$\\
Feb, $24$ & $49.1$  & Oct, $08$ & $18.0$\\
May, $31$ & $46.8$  & Set,$ 11$ & $18.4$\\
May, $21$ & $45.5$  & Apr, $15$ & $18.4$\\
May, $14$ & $44.6$  & Oct, $13$ & $18.5$\\
Jun, $23$ & $44.6$  & Apr, $17$ & $18.5$\\
Jul, $13$ & $44.3$  & Nov, $10$ & $18.8$\\
Jun, $24$ & $42.8$  & Nov, $09$ & $18.9$\\
Jul, $12$ & $42.7$  & Mar, $06$ & $19.1$\\
Jan, $21$ & $41.5$  & Oct, $12$ & $19.2$\\
Jul, $29$ & $41.4$  & Mar, $17$ & $19.3$\\
Jun, $15$ & $41.2$  & Feb, $28$ & $19.5$\\
Jun, $27$ & $40.5$  & Oct, $16$ & $19.5$\\
Mar, $20$ & $40.5$  & Apr, $13$ & $19.5$\\
Mar, $11$ & $39.9$  & Nov, $26$ & $19.5$\\
Aug, $22$ & $39.7$  & Set, $09$ & $19.6$\\
Jan, $25$ & $39.5$  & Set, $20$ & $19.7$\\
\hline
\hline
\end{tabular}
\end{center}
\caption{Ranking of the 20 worst/best days of the year $2010$ according to their daily average delay for flights with positive delay.}
\label{tab:DayRanking}
\end{table}
 
\begin{figure}[!h]
\begin{center}
\includegraphics[angle=-90,width=8cm]{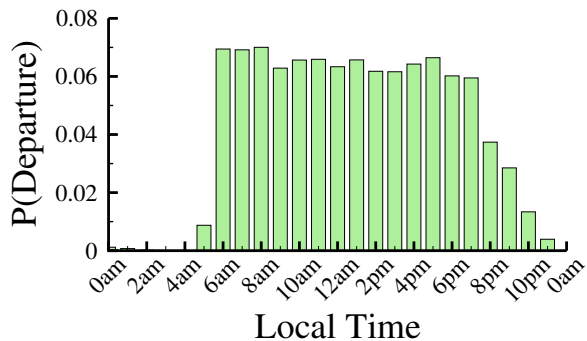}
\caption{Probability of flight departure as a function of the scheduled departure hour.}
\label{fig:ProbFlDep}
\end{center}
\end{figure}

The United States spans through several time zones. In order to unify criteria and simplify the analysis, we transform all the local times to the East Coast local time. Olson or tz database~\cite{tz} is used to ensure an accurately timezones conversion from the respective local times in the database to the East Coast local time (EST in winter and EDT in summer time). Plotting the departure probability as a function of the scheduled departure  in Figure~\ref{fig:ProbFlDep}, we can distinguish a zone that goes from 00am to $4$am with almost no operations. We set thus the start of a new day in our analysis and simulations at $4$am East Coast local time, that is, at $3$am Central, $2$am Mountain and $1$am Pacific. In this way, the starting of the new day coincides with the low activity phase of air operations in most of the country.

We also notice that for daily networks $98\%$ of the edges are bidirectional on average, i.e., if there is a flight from A to B there is always a flight from B to A. Taking this into account, we symmetrized the network to simplify the cluster analysis.

\subsection{Annual fraction of connecting passengers for each US commercial airport}
\label{sec:DataConPass}
Another key input for modeling the delay propagation over the network is the connection between flights. The previous database has no information regarding flight connectivity, neither for the crews nor the passengers. In order to at least approximate the heterogeneity of the airports in this sense, we used the T100 Domestic Market (US carriers) and the DB1B Ticket information downloaded from the BTS page~\cite{note2}. These documents allow us to obtain an approximation of the annual fraction of connecting passengers for each airport. The information of T100 corresponds to the total 
number of passengers who have a flight departing from an airport regardless of their real point of origin ($Passengers_{T100}$). On the other hand, the database DB1B contains a $10\%$ sample of the 
number of passengers whose itinerary originated in each given airport ($Passengers_{DB1B}$). So for each airport we can get an approximation of the annual fraction of connecting passengers as:

\begin{equation}
\frac{Passengers_{T100}-10.Passengers_{DB1B}}{Passengers_{T100}}
\end{equation}

Although our model is based on flight not passenger connectivity, we assume that these ratios are related, which is always better than assuming arbitrary values, with $\alpha$ controlling the intensity of such relation. 

\begin{table}[!h]
\begin{center}
\begin{tabular}{c|c}
Airport code & Fraction of connecting passengers \\
\hline
\hline
ATL & $0.81$\\
ORD & $0.72$\\
DFW & $0.75$\\
DTW & $0.69$\\
DEN & $0.71$\\
MSP & $0.68$\\
IAH & $0.75$\\
SLC & $0.73$\\
MEM & $0.81$\\
MCO & $0.69$\\
\hline
\hline
\end{tabular}
\end{center}
\caption{Fraction of connecting passengers for the top ten airports in degree.}
\label{tab:AirportFracConn}
\end{table}

\section{Model description}
To simulate the delay propagation, we developed an agent-based model that combines within the same framework queuing and a
schedule based approach dynamics.

\subsection{Overview}
\label{sec:ModelOverview}
As stated in the main text, one of the purposes of this model is to understand how delays propagate and magnify considering internal operational factors and schedule. As it will be explained further below, "extrinsic" or primary delay is given at the initial steps of the simulation to the first flight of the day for some aircraft rotations, and then let this perturbation evolve multiplying or diminishing the delay according to the particular structure of the system. Concretely, the model dynamics will be based on three subprocesses which are: (i) aircraft rotation, (ii) flight connectivity and (iii) airport congestion. The last two are independent from each other, and can be turned on/off to explore the relevance of each subprocess in the delay propagation dynamics.  Aircraft rotation, on the other hand, is intrinsic to the schedule and so we do not switch it off.

We use one-minute intervals as the basic time step unit in the model and proceed in each simulation until the schedule of a selected day is completed (all flights had completed their itinerary). In most cases, this means slightly more than $1,440$ minutes. This time interval allows the simulation to execute actions at a realistic concurrent time-scale and is the finest level available in the data. As 
mentioned in the previous section, $4$am East Coast local time is set as the starting point for airport operations and to begin the aircraft rotational sequences. By this selection, we ensure that most aircraft rotational sequences are sorted correctly and it is the natural choice considering the daylight time flow in the United States. Also, as mentioned before, to arrange the schedule in a real sequential order we converted time data from Local operation time to Eastern local time.

\subsection{Hierarchy of Objects}

\subsubsection{Aircraft (tail-number)}
\label{sec:ModelAircraft}
The airplane is the primary fundamental agent of the simulation. The number of airplanes that participate in the simulation varies with the day considered, but it is around $4,000$.  Each 
aircraft is unique and comes identified by their tail number. This code allows us to reconstruct the rotational sequence of the plane during the day. This sequence can be subdivided in individual flight legs or point-to-point flights.  

\subsubsection{Point-to-point flight}
\label{sec:ModelFlight}
\begin{figure}[h]
\begin{center}
\includegraphics[width=8.6cm]{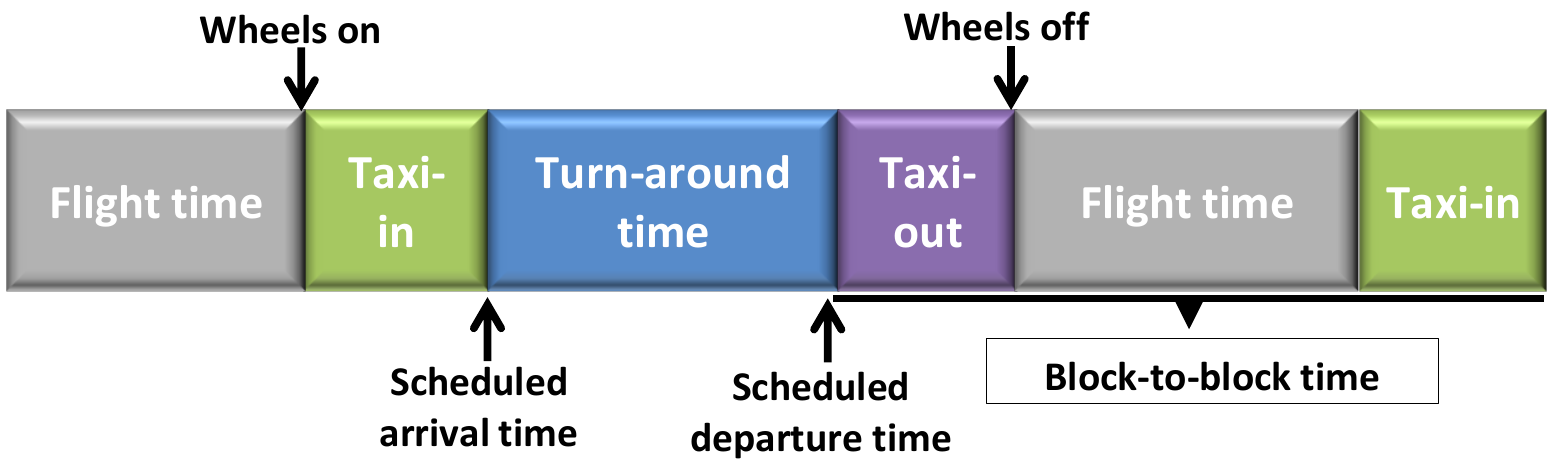}
\caption{Turn-around and block-to-block time/phase definition.}
\label{fig:GenericTimeline}
\end{center}
\end{figure}

This is the basic schedule unit. It is the minimum package of information used as an input to relocate an aircraft from an origin to a destination airport, meeting the planned schedule. 
During their itinerary an aircraft can be in one of two flight phases: block-to-block or turn-around phase. The former is the time elapsed from the airport origin gate to the airport destination 
gate. The latter is defined as the time the aircraft remains parked at the airport gate (Figure~\ref{fig:GenericTimeline}).
 
Flights are characterized by a tail-number, origin airport, destination airport, schedule departure time ($T_{sch.d}$) and schedule arrival time ($T_{sch.a}$). Block-to-block time ($T_{b}$) 
between two airports is calculated as:
\begin{equation}
T_{b}=T^{j}_{sch.a}-T^{i}_{sch.d} ,
\end{equation}
where j corresponds to destination airport and i to the origin one.  Another issue worth noting in our model is that, in the block-to-block phase we do not allow for delay absorption or reduction. This could only be achieved in the turn-around phase by means of the difference between the actual arrival time of the previous flight leg and the scheduled departure time of the next flight leg.

\subsubsection{Air carrier (airline id)}
\label{sec:ModelAirline}
Air carriers are the second level unit in the model. Each aircraft has an airline associated via the airline code id. Only aircrafts having the same airline id are allowed to interact during the process of flight connectivity (see \ref{sec:ModelFlConn} for further details). 

\subsubsection{Airport}
\label{sec:ModelAirport}
The airport is an intermediate-level entity located in space coordinates, where interactions among aircrafts take place. This interaction occurs indirectly through the schedule, flight connections or airport queues (see \ref{sec:ModelAirportCong} for further details). Each airport is different from the others because of their planned capacity and the local aggregation of the schedule. Airports play the role of nodes in the transport network.  

\subsubsection{Clusters of congested airports}
\label{sec:ModelCluster}
This is a high-level entity that represents interactions between airports. The clusters are formed by airports whose average (departure) delay per flight is higher or equal to $29$ minutes and 
are linked by a direct connection (see \ref{sec:ModelClustering} for further details). In most cases, we are interested in the largest cluster of the full day (or by hour in some cases). The size of a cluster is measured according to the number of airports that belong to it. 

\begin{figure}[!h]
\begin{center}
\includegraphics[width=8.6cm]{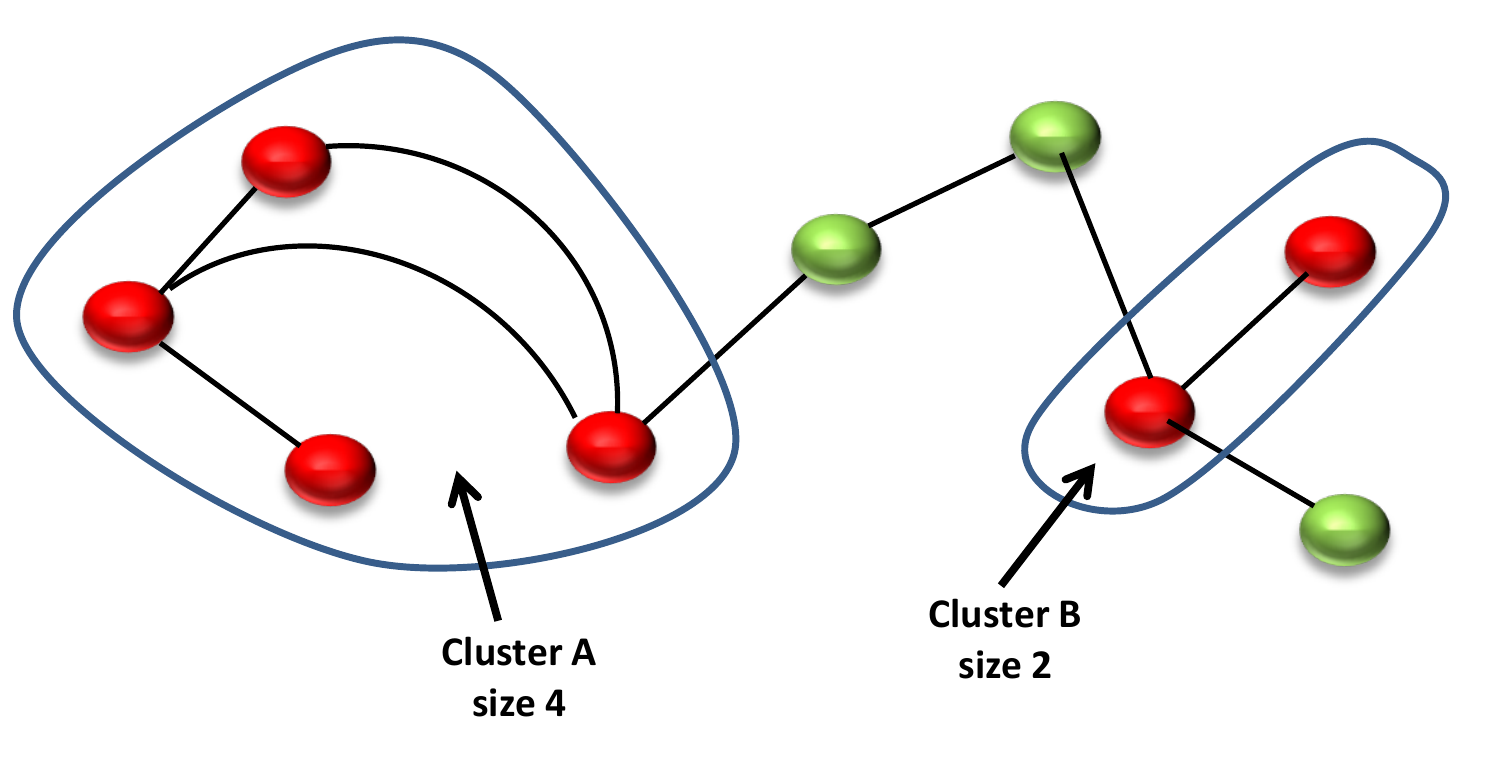}
\caption{Red dots correspond to airports whose average departure delay per flight in a certain time period is $>$ than 29 minutes. Green dots correspond to airports whose average departure 
delay per flight in a certain time period is $leq$ than 29 minutes.}
\label{fig:ClusterEx}
\end{center}
\end{figure}

Figure~\ref{fig:ClusterEx} shows a representation of two clusters (Cluster A and B) constituted by airports whose average departure delay per flight in a certain time period is equal or larger than $29$ minutes (red dots). Apart from this condition airports within these clusters are linked by a direct connection. In this case, cluster A correspond to the largest cluster in a certain time period according to the number of airports that form this cluster.

\subsection{Subprocesses}

\subsubsection{Aircraft rotation}
\label{sec:ModelAircraftRot}
\begin{figure}[h]
\begin{center}
\includegraphics[width=8.6cm]{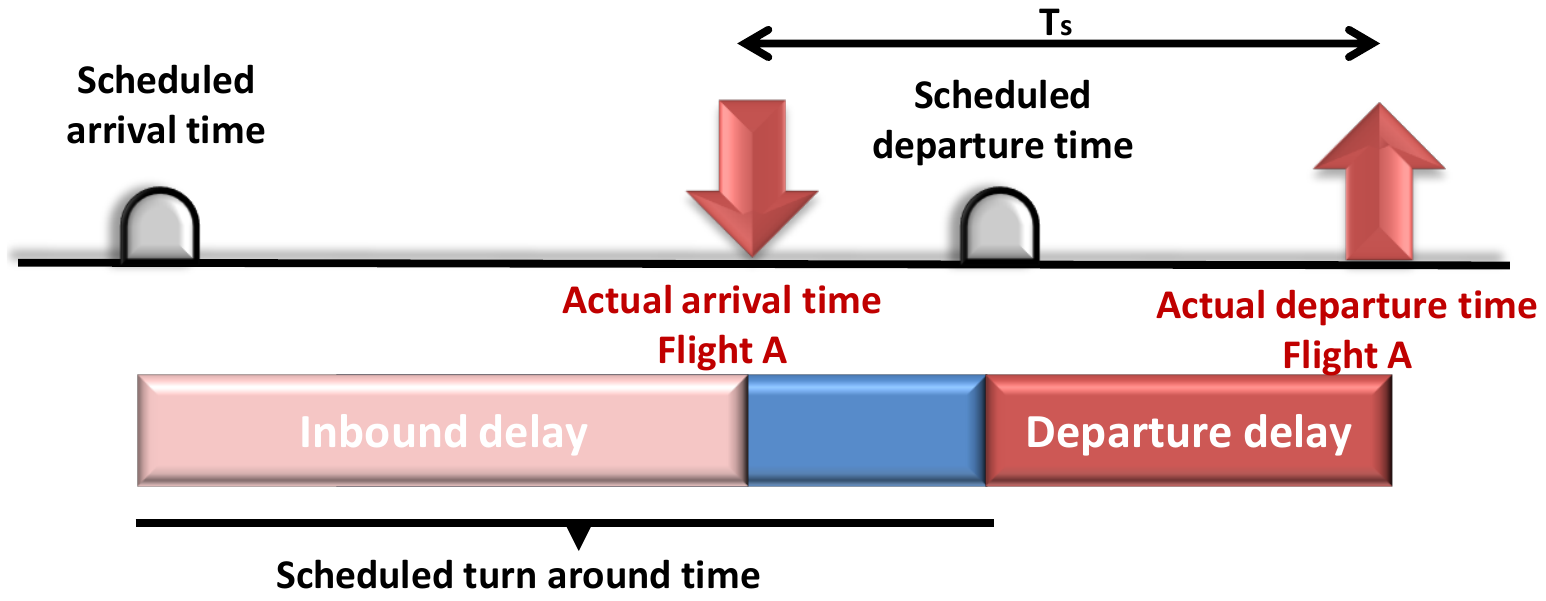}
\caption{Aircraft rotation description.}
\label{fig:RotationalTimeline}
\end{center}
\end{figure}

During a day, each aircraft has an itinerary to accomplish that in the vast majority of cases consists of two or more flight legs. Naturally, to complete a flight leg, the previous ones have to be fulfilled, e.g., it is not possible to depart from San Francisco to Honolulu if the airplane has not completed the previous leg from Atlanta to San Francisco.  Besides this evident situation, if 
an aircraft arrives late (inbound delay) and the delay cannot be absorbed by the turn-around time it will depart late in the next flight leg (Figure~\ref{fig:RotationalTimeline}). Usually, a buffer time is included in the turn-around phase to absorb this type of delay but this is already incorporated in the schedule obtained from the data.

Another feature of this subprocess, is that in the turn-around phase each aircraft, when arrived, has to comply with a minimum service time $T_{s}$, in the simulations set as 30 minutes. This service time includes operations such as refueling, passenger unboarding/boarding, luggage handling, safety inspection, etc.

\subsubsection{Flight connectivity}
\label{sec:ModelFlConn}
In addition to rotational reactionary delay, the need to wait for load, connecting passengers and/or crew from another delayed airplane from the same fleet (airline id) may cause, as well, reactionary delay.

\begin{figure}[h]
\begin{center}
\includegraphics[width=8cm]{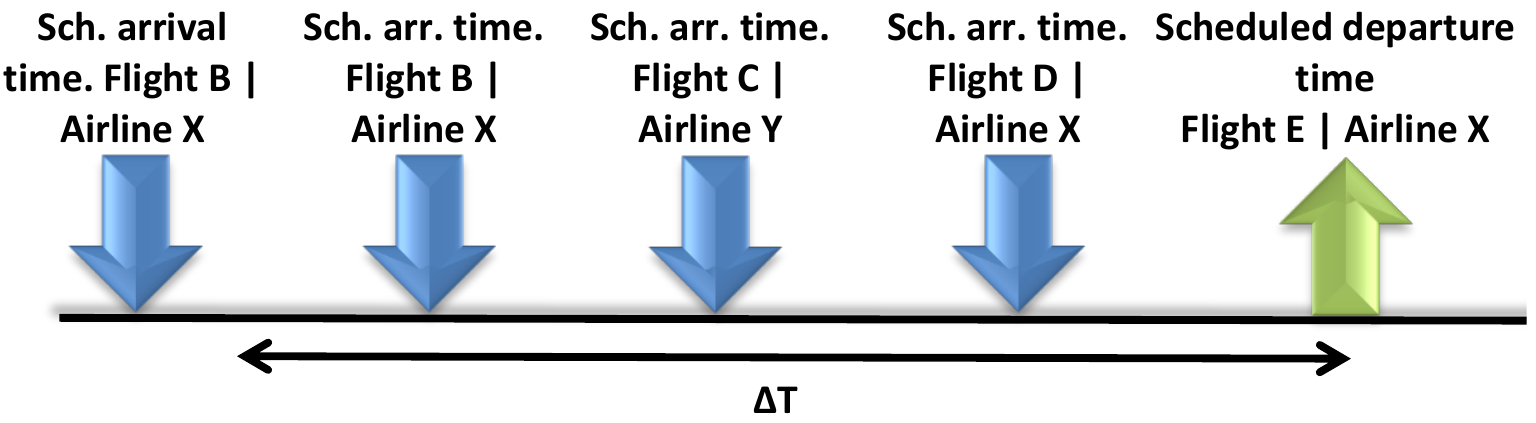}
\caption{Possible connections within flights of the same airline.}
\label{fig:PossibleConn}
\end{center}
\end{figure}

For each flight at a particular airport, connections from that airport are randomly chosen as follows. Firstly, we take a $\Delta T$ window prior to the scheduled departure time of the flight. Secondly, we distinguish possible connections of the same airline from other flights, that have a scheduled arrival time within the $\Delta T$ window (Flights B and D in the example of 
Figure ~\ref{fig:PossibleConn}). Finally, from these possible connections we select those with probability $\alpha * \textrm{flight connectivity factor}$. The flight connectivity factor was defined in ~\ref{sec:DataConPass} and $\alpha$ is an effective parameter of control that allows to modify the strength of this effect in the model.  For instance, $\alpha=0$ means that there is no connection between flights with different tail number, while $\alpha=1$ makes the fraction of connecting flights of the same airline equal to the fraction of connecting passengers in the given airport. In the simulations, $\alpha$ is varied according to the case under study and $\Delta T$ is always taken to be $180$ minutes ($3$ hours).

Let us suppose that from the previous example Flight D was randomly selected.

\begin{figure}[h]
\begin{center}
\includegraphics[width=8cm]{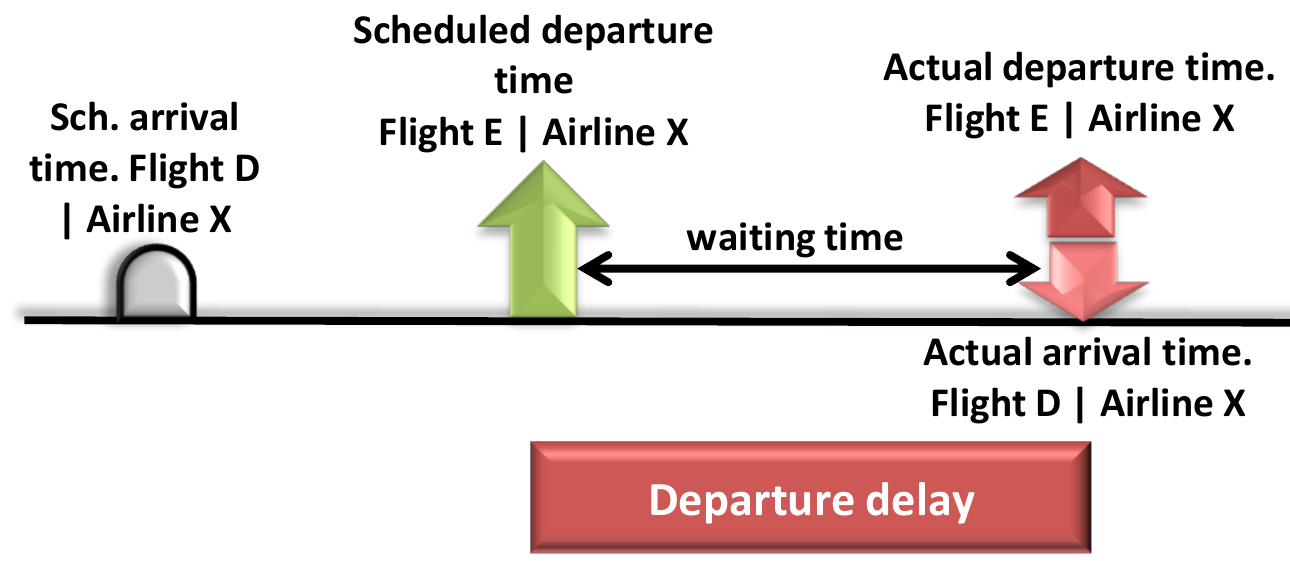}
\caption{Flight connectivity description.}
\label{fig:FlightConn}
\end{center}
\end{figure}
\noindent
By this subprocess an airplane is able to fly if and if only their connections have already arrived to the airport, if not it has to wait until this condition is satisfied (Figure ~\ref{fig:FlightConn}). It is important to note that flight connectivity is the only source of stochasticity in the model due to a lack of knowledge about the real flight connections within the schedule.

\subsubsection{Airport congestion}
\label{sec:ModelAirportCong}
Because airports are entities with a finite capacity, the possibility of their congestion has to be introduced in the model. Interactions between aircrafts other than the ones defined by the schedule (flight connectivity and aircraft rotation) are in this way taken into account. This occurs indirectly through an airport's queue. That is to say that delays from airplanes of different airlines can delay others because they congest the airport. The delay spreading does not surge so easily as in the previous cases, it requires a cumulative effect of several delayed aircrafts to perturb the airport efficiency and once this condition is meet the delay spread to other aircrafts and affect other airlines. 

\begin{figure}[h]
\begin{center}
\includegraphics[angle=-90,width=8cm]{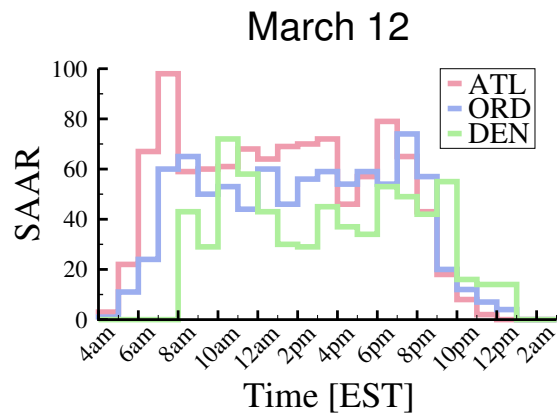}
\caption{Example of SAAR for three major airports: Atlanta International Airport (ATL), O'Hare International Airport (ORD) and Denver International Airport (DEN).}
\label{fig:SAAR}
\end{center}
\end{figure}

We assume a "First in-First Served" queuing protocol that is the most widely used queue operation and simple to introduce in the model.
In the simulations each airport will have a capacity that varies throughout the day according to the Scheduled Airport Arrival Rate ($SAAR$). This means that for every airport we measure the scheduled flights that arrive per hour and this is the nominal capacity for each hour of the day (Figure~\ref{fig:SAAR}). Due to reactionary delays aircrafts may not arrive as planned and the Real Airport Arrival Rate ($RAAR$) will vary. Whenever $RAAR>SAAR$, a queue begins to form with the arriving aircrafts. Naturally, airplanes that are not in queue are being served and this service time lasts $T_{s}$ (see ~\ref{sec:ModelAircraftRot} for further details). It should be noticed that once an aircraft starts to be served this process cannot be interrupted no matter how $SAAR$ varies.
We define another effective control parameter $\beta$ in order to modify the nominal capacity of the airports. This parameter multiplies the $SAAR$ and in the simulations presented here affects all the airports in the same way. For instance, if we want to introduce a buffer capacity of $20\%$, $\beta$ is set to $1.2$.

\subsection{Initial conditions}
\label{sec:ModelInitCond}
Initial condition refers to the situation of the first flight of an aircraft sequence, meaning when, where and the departure delay of this flight. As mentioned in the main text, variations on this situation can have a great impact on the delay propagation. In other words, the dynamics of delays over the network is highly sensitive to the initial conditions.

\begin{figure}[ht]
\begin{center}
\includegraphics[width=8cm]{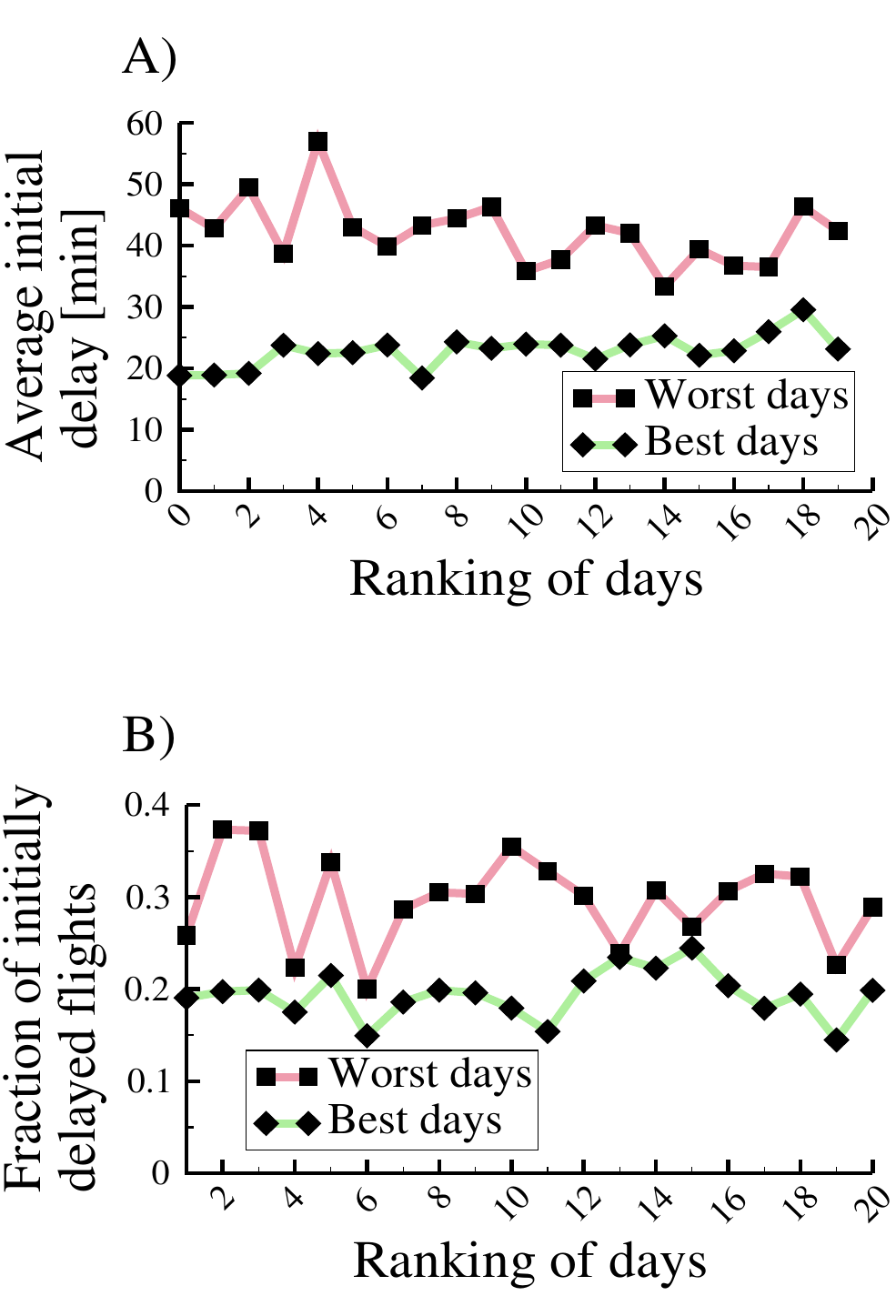} 
\caption{Initial conditions of the 20 worst days (red) and of the 20 best days (green) of 2010.}
\label{fig:InitialCond}
\end{center}
\end{figure}

\noindent
We characterized initial conditions by the average delay per flight for the first flights of all the aircraft sequences and by the fraction of airplanes that their first flight was delayed. Comparing the ranking of the $20$ worst and best days of $2010$ (Figure~\ref{fig:InitialCond}) we can observe that it is most likely that if a day started with unfavorable initial conditions it 
will likely produce large congested clusters.

The simulations can be initialized by two different ways depending on the case under study: from data or random initial conditions.
\subsubsection{From the data}
\label{sec:ModelInitCondData}
Initializing the model "from the data" means to replicate exactly the situation of the first flights of all the aircrafts sequences for a particular day.

\subsubsection{Random initial conditions}
\label{sec:ModelInitRandom}
When random initial conditions are set, initial delays are reshuffled among all possible aircrafts, so when and where may vary. Two inputs are needed: initial delay and fraction of flights 
initially delayed. For instance,

Initial delay: $20$ minutes

Percentage of airplanes initially delayed: $10\%$

Suppose that the number of aircrafts for one day simulation is $4,000$. 
In this example, $400$ aircrafts will have their first flight departing with an initial delay of 20 minutes.

\subsection{Decision Tree}

Model flowchart summary including all the subprocesses. Flowchart objects in green and red will be explained separately.

\label{sec:ModelTree}
\begin{figure}[ht]
\begin{center}
\includegraphics[width=8cm]{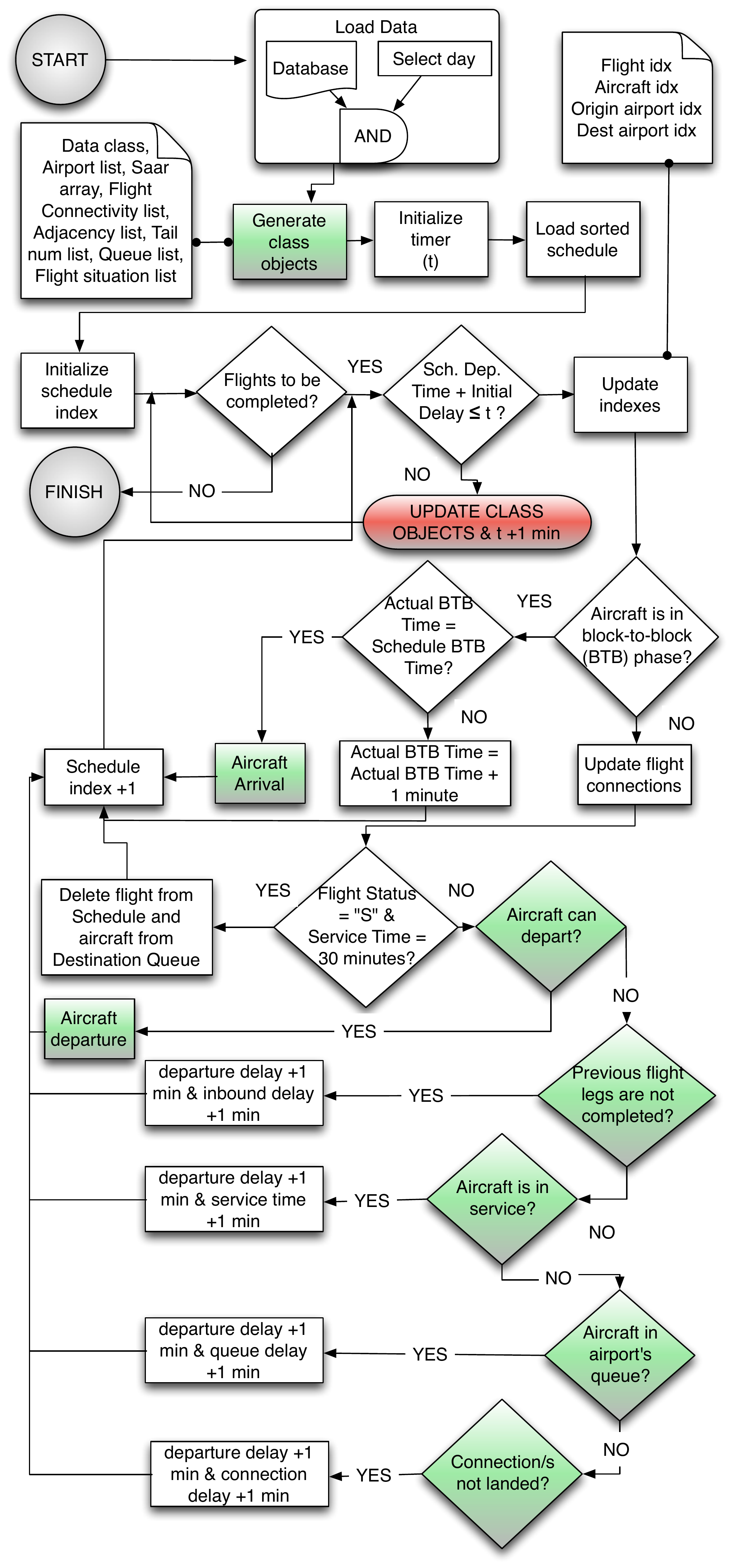} 
\label{fig:ModelFlow}
\end{center}
\end{figure}

\begin{itemize}
 \item \emph{Generate class objects}: \newline
Once the data is loaded for a particular day into the data class object, the remaining class objects are created using this data structure.  These objects are:
\begin{itemize}
 \item Airport list: Indexation of all the airports that operated that day.
 \item SAAR matrix: Includes the hourly capacity (schedule airport arrival rate) for every airport in the list.
 \item Airport Flight Connectivity Factor List.
 \item Adjacency list: Contains the network structure for that day.
 \item Tail number: Indexation of all the aircrafts that operated that day. 
 \item Schedule: For each flight the schedule object contains the information described in section~\ref{sec:ModelFlight}, initial delay (see ~\ref{sec:ModelInitCond}), flight index, 
flight status ( on land "L", flying "F" and in service or in queue "S"), inbound index (previous flight leg index) and connections (see ~\ref{sec:ModelFlConn}). All flights are initialized
with flight status "L".
 \item Tail number situation: For each aircraft contains the origin airport, the destination airport, the scheduled and actual block-to-block time and the departure delay (initial, inbound, queue 
and due to connections).
 \item Airports tail number queue: For each airport contains the aircrafts ordered as First in - First served.
 \item Airports flight queue: The same as the previous one but indexed with flight number.
\end{itemize}
\item \emph{Update class objects \& t +1 min}: \newline
Objects as Schedule and Airport tail number and flight queues are synchronous updated for each time step.
\item \emph{Aircraft arrival}: \newline
The flight status is changed from "F" to "S" and the airport's tail number and flight queues are updated.
\item \emph{Aircraft can depart?}: \newline
The aircraft can depart if the service time (30 minutes) is complete and the are no flight connections to wait for. Initial flight legs of an itinerary are considered as already served.
\item \emph{Aircraft departure}: \newline
 Tail number situation and  origin airport queues are updated. The actual block-to-block time is reset. Flight status is changed from "L" to "F".
\item \emph{Previous flight legs are not completed?}: \newline
Check if the inbound index is among the flight connections and the flight status is "L".
\item \emph{Aircraft is in service?}: \newline
Inspect if the flight status for the aircraft is "S" and the service time is different from zero or the aircraft position at the airport queue is less than airport capacity.
\item \emph{Aircraft is in airport's queue?}: \newline
Check if the flight status is "S" and the service time is zero.
\item \emph{Connection/s not landed?}: \newline
Verifies if the number of connections in the schedule for the flight is zero and the flight status is "L".
\end{itemize}

\subsection{Clustering}
\label{sec:ModelClustering}
\begin{enumerate}
\item Create a cluster list with all airports labeled as -1 (unexplored).
\item Create an empty list (active list) to include the airports to inspect while traversing the adjacency list (network).
\item While unexplored airports continue to exist in the cluster list:
 \begin{itemize}
 \item For each airport in the cluster list:
  \begin{itemize}
    \item Check if the airport is unexplored and the average delay per flight for the airport is greater than 29 minutes.
    \item If it is so, label the airport with its index and insert the airport index in the active list.
    \item Else, label the airport as -2 (not delayed).
    \item While the active list continue to have airports to explore:
    \begin{itemize}
    \renewcommand{\labelitemiii}{$\circ$}
    \item For each airport in the active list:
    \begin{itemize}
    \renewcommand{\labelitemiv}{$\ast$}
      \item Explore its neighbors in the adjacency list.
      \item Check if they are labeled as unexplored and their average delay per flight is greater than 29 minutes.
      \item If it is so, label them with the same index as before and insert the airport index in the active list.
      \item Else, label the airport as "not delayed".
    \end{itemize}
    \item Remove from the active list the airports that  their neighbors had been explored.
 \end{itemize} 
 \end{itemize}
 \end{itemize}
 \end{enumerate}

\subsection{Overview of the model parameters}
\label{sec:ModelParameters}

\begin{table}[h]
\begin{center}
\begin{tabular}{c|c|c|c|c}
Parameter & October 27 & March 12 & December 12 & July 13 \\
\hline
\hline
$T_{s}$ [min] & \multicolumn{4}{c}{30} \\ \hline
$\Delta T$ [min] & \multicolumn{4}{c}{180}\\ \hline
$\alpha$ & 0.263 & 0.190 & 0.265 & 0.075\\ \hline
$\beta$ & \multicolumn{4}{c}{1.0} \\ \hline
Initial Condition & \multicolumn{4}{c}{"From the data"}  \\
\hline
\hline
\end{tabular}

\begin{tabular}{c|c|c}
Parameter &  October 9 & April 19 \\
\hline
\hline
$T_{s}$ [min] & \multicolumn{2}{c}{30} \\ \hline
$\Delta T$ [min] & \multicolumn{2}{c}{180}\\ \hline
$\alpha$ & 0.020 & 0.020\\ \hline
$\beta$ & \multicolumn{2}{c}{1.0} \\ \hline
Initial Condition & \multicolumn{2}{c}{"From the data"}  \\
\hline
\hline
\end{tabular}
\end{center}
\caption{Overview of default values of the model's parameters. The values of $\alpha$ correspond to the best fit for the day.}
\label{tab:ModelParameters}
\end{table}

\section{Model simulations}
\label{sec:ModelValidation}

\begin{figure}[ht]
\begin{center}
\includegraphics[width=8cm]{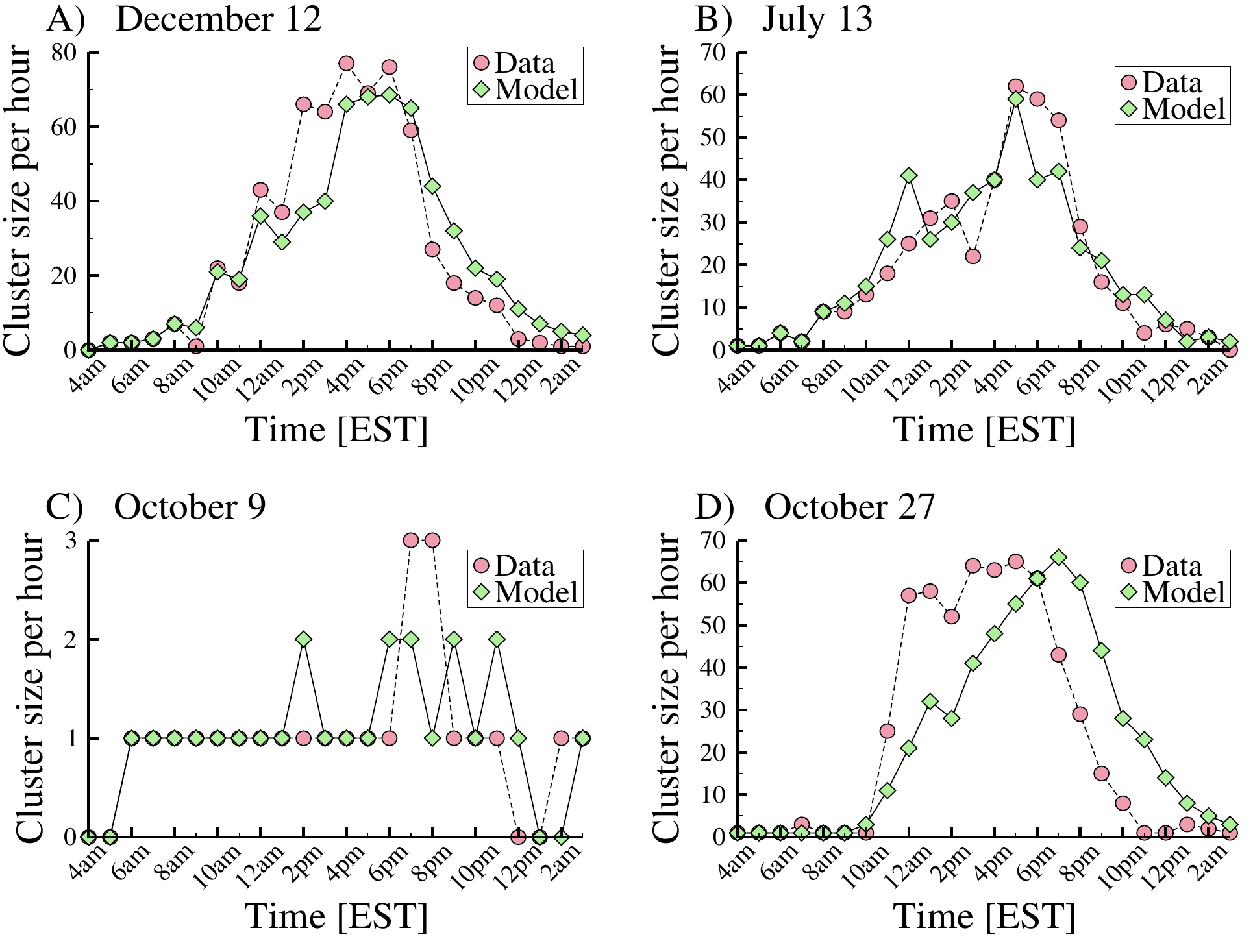} 
\caption{Evolution of the largest cluster size for A) December 12 ($\alpha$ = 0.265), B) July 13 ($\alpha$ = 0.075), C) October 9 ($\alpha$ = 0.002) and D) October 27 ($\alpha$ = 0.263)}
\label{fig:ClusterSize}
\end{center}
\end{figure}

\subsection{Model validation and sensitivity to $\alpha$}

\begin{figure}[ht!]
\begin{center}
\includegraphics[width=8cm]{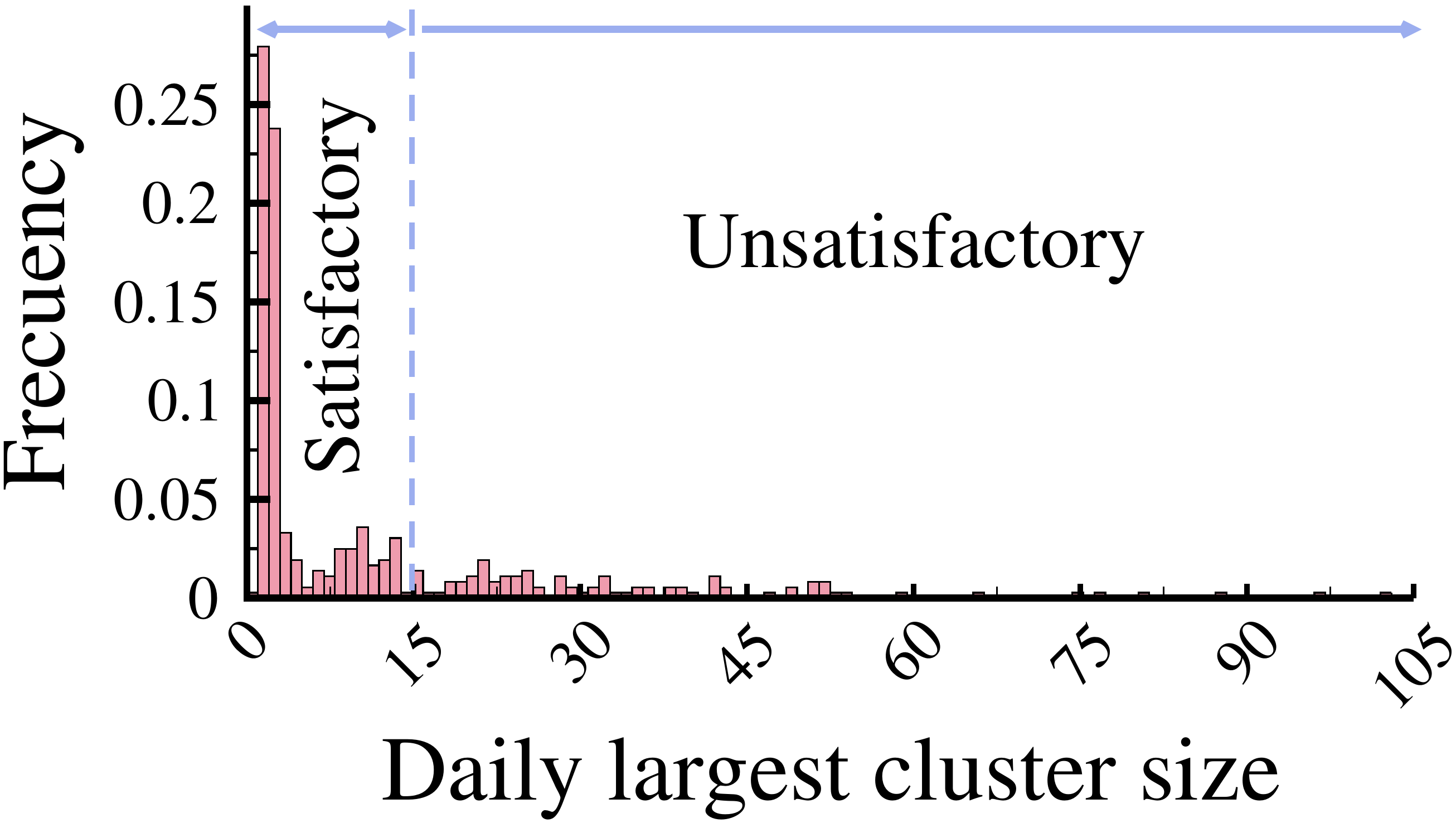}
\caption{Frequency of the largest cluster size for all days of $2010$.}
\label{fig:HistDBC}
\end{center}
\end{figure}

\begin{figure}[hb!]
\begin{center}
\includegraphics[angle=-90,width=8cm]{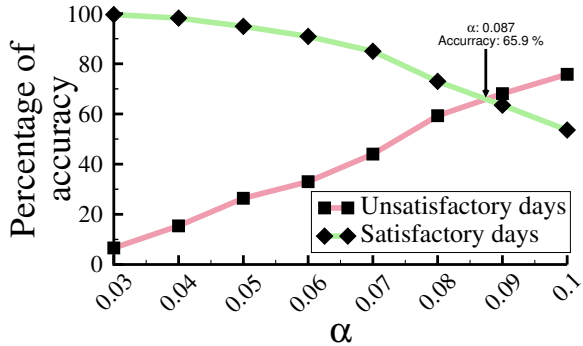}
\caption{Exploring the model forecast accuracy by varying the $\alpha$ parameter. All days of 2010 are taken into account.}
\label{fig:success_ratio}
\end{center}
\end{figure}

Figure~\ref{fig:ClusterSize} displays results for other days different from the ones presented in the text. Results for December $12$ (Figure~\ref{fig:ClusterSize} A), July $13$ (Figure~\ref{fig:ClusterSize} B) and October $9$ (Figure~\ref{fig:ClusterSize} C) confirm that the model is in good agreement with the data when $\alpha$ is fitted for each day. In the case of October $27$ (Figure~\ref{fig:ClusterSize} D), the size of the cluster evolved much faster than the model prediction, although the size could be predicted. Analyzing the possible explanation to this difference, we found that severe weather conditions occurred that day across an important part of the country~\cite{note} affecting flights in airports such as Hartsfield-Jackson (Atlanta), John F. Kennedy (New York), La Guardia (New York), St. Paul (Minneapolis), O'Hare (Chicago), Philadelphia and Newark. External perturbations were not explicitly introduced in the model so we cannot expect to be able to reproduce well these days delay dynamics.

In the previous sections we have defined days/airports with problems as those whose average delay per delayed flight was over $29$ minutes. Another way, related to the previous one, of classifying the days is by means of the largest cluster size of the day. To do so, we set a cluster size that corresponds to $15$ airports so that if the largest cluster size in a day is higher than this threshold the day is labeled as problematic or unsatisfactory. On the other hand, if it is less than $15$ airports the day is labeled as satisfactory. This threshold was selected because in the distribution of largest cluster size there exists a small depression at this value (Figure~\ref{fig:HistDBC}). This particular value for the threshold is arbitrary. Still, we have repeated the analysis with some other thresholds and checked that the main conclusions are maintained.

\begin{table}[h]
\begin{center}
\begin{tabular}{c|c||c|c}
\multicolumn{2}{c}{Unsatisfactory days} & \multicolumn{2}{c}{Satisfactory days} \\
\hline
 DATE & Accurate Prediction & DATE & Accurate Prediction \\
\hline
\hline
Oct, $27$ & No  & Apr, $19$ & Yes\\
Mar, $12$ & Yes & Oct, $09$ & Yes\\
Dec, $12$ & Yes & Nov, $11$ & Yes \\
Jan, $24$ & No  & Apr, $14$ & Yes\\
Feb, $24$ & Yes & Oct, $08$ & No\\
May, $31$ & No  & Set, $11$ & Yes\\
May, $21$ & Yes & Apr, $15$ & Yes\\
May, $14$ & No  & Oct, $13$ & Yes\\
Jun, $23$ & Yes & Apr, $17$ & Yes\\
Jul, $13$ & Yes & Nov, $10$ & Yes\\
Jun, $24$ & No  & Nov, $09$ & Yes\\
Jul, $12$ & Yes & Mar, $06$ & Yes\\
Jan, $21$ & Yes & Oct, $12$ & Yes\\
Jul, $29$ & Yes & Mar, $17$ & No\\
Jun, $15$ & Yes & Feb, $28$ & Yes\\
Jun, $27$ & No  & Oct, $16$ & Yes\\
Mar, $20$ & Yes & Apr, $13$ & Yes\\
Mar, $11$ & Yes & Nov, $26$ & Yes\\
Aug, $22$ & Yes & Set, $09$ & No\\
Jan, $25$ & Yes & Set, $20$ & Yes\\
\hline
\hline
\end{tabular}
\end{center}
\caption{Ranking for the top 20 days by the average delay for flights with positive delay. Model accuracy according to the classification of each day in satisfactory or unsatisfactory. The model is 
able to predict unsatisfactory days with an accuracy of 70\% and satisfactory ones with an 85\%.}
\label{tab:Accuracy}
\end{table}

\begin{table}[!h]
\begin{center}
\begin{tabular}{c|c|c}
\multicolumn{3}{c}{March 12} \\
\hline
Airport Code & Precentage Realizations & Accurate Prediction \\
\hline
\hline
ATL & 100.0 & Yes\\
CWA & 100.0 & Yes\\
DFW & 100.0 & No\\
DLH & 100.0 & Yes\\
EAU & 100.0 & No\\
EYW & 100.0 & Yes\\
FLL & 100.0 & Yes\\
GGG & 100.0 & No\\
MGM & 100.0 & Yes\\
MIA & 100.0 & Yes\\
ORD & 100.0 & No\\
SJT & 100.0 & No\\
STT & 100.0 & Yes\\
TOL & 100.0 & No\\
BHM & 99.8  & Yes\\
CAK & 99.5  & Yes\\
CHA & 98.6  & Yes\\
FAY & 98.4  & Yes\\
MEM & 98.3  & Yes\\
HSV & 97.9  & No\\
\hline
\hline
\end{tabular}
\end{center}
\caption{Top 20 ranking of airports that appear more frequently in the largest cluster for the model results compared to what actually occured on March 12.}
\label{tab:Accuracy2}
\end{table}

The introduction of the threshold allow us to define a binary variable associated to the performance of the network each day. Since the model requires a fit in $\alpha$ to reproduce the precise dynamics of the congested clusters, the aim of this exercise is to set a generic value of $\alpha$ and study how many of the satisfactory/unsatisfactory days are actually predicted. According to our definition, during $2010$, $75\%$ of the days get a satisfactory performance. In order to assess the model correspondence with reality, we have to take into account that satisfactory days outweigh unsatisfactory ones. Naturally, with a high $\alpha$ the model simulations predict unsatisfactory days with high accuracy but provide many false positives for satisfactory days. On the other hand, with a low $\alpha$, most of days with small clusters are successfully predicted but not those with large congested clusters. Bearing this in mind, we defined the percentage of accuracy as a tradeoff between the percentage of accuracy for satisfactory and unsatisfactory days. Figure~\ref{fig:success_ratio} show the fraction of correct predictions both for satisfactory and unsatisfactory days. Both curves cross at a value of $\alpha = 0.087$ and at an accuracy rate of $65.9\%$. Obviously, this is a simplistic technique to measure performance. A more elaborate technique should include appropriate economic considerations to take into account that the cost related to false positives, claiming that a day is going to have a large congested cluster without actually occurring, and false negatives, not being able to predict a major collapse, are different. Even so, this simple method provides us with a quantitative framework to validate the model and to assess the importance of including further mechanisms in the simulation.

Another accuracy test was done to check if the model is able to predict not only the size but the airports that comprises the largest cluster of the day. We selected March $12$ whose largest cluster is formed by $97$ airports. The model is stochastic, so we run it for $1500$ realizations. Comparing the data with the model results for the top $97$ airports most frequently appearing in the largest cluster, the model accurately identify $57.8\%$ of them. Table~\ref{tab:Accuracy2} displays the Top $20$ airports which are more prevalent in the simulations showing if they appeared in the real data as part of the largest cluster for March $12$. This is a first comparison, since the real cluster is coming from a single realization in a particular day it cannot be taken as a definitive validation of the model. However, an accuracy of $57.8 \%$ with such a simple framework is already encouraging.

\subsection{Analysis of the model sensitivity to changes in $\beta$}
\label{sec:ModelSensitivity}

\begin{table}[h!]
\begin{center}
\begin{tabular}{c|c||c|c}
\multicolumn{2}{c}{April, 19} & \multicolumn{2}{c}{March, 12} \\
\hline
 $\beta$ & Average largest & $\beta$ & Average largest\\
 $\,$ & cluster size [airports] & $\,$ & cluster size [airports]\\
\hline
\hline
$1.0$ & $1.0$  & $1.0$ & $92.0$\\
$0.9$ & $1.0$  & $1.1$ & $92.3$\\
$0.8$ & $1.0$  & $1.2$ & $89.9$\\
$0.7$ & $1.0$  & $1.3$ & $88.7$\\
$0.6$ & $1.3$  & $1.4$ & $86.4$\\
$0.5$ & $16.8$ & $1.5$ & $86.4$\\
\hline
\hline
\end{tabular}
\end{center}
\caption{$\beta$ variation for April 19 with an $\alpha$ fixed at 0.02: airports should work at half the scheduled capacity to transform this day into an unsatisfactory one. $\beta$ variation for 
March 12 with an $\alpha$ fixed at 0.19: an increase of 50\% in $\beta$ decreases the size of the cluster by only a 7\%.}
\label{tab:beta}
\end{table}

As stated in the main text, the model is able to reproduce the clusters of congested airports by fitting $\alpha$ while fixing $\beta$ to $1$. In fact, as shown in Table~\ref{tab:beta} the model 
has a low sensitivity to a variation in the $\beta$ coefficient.

\begin{figure}[h!]
\begin{center}
\includegraphics[width=8cm]{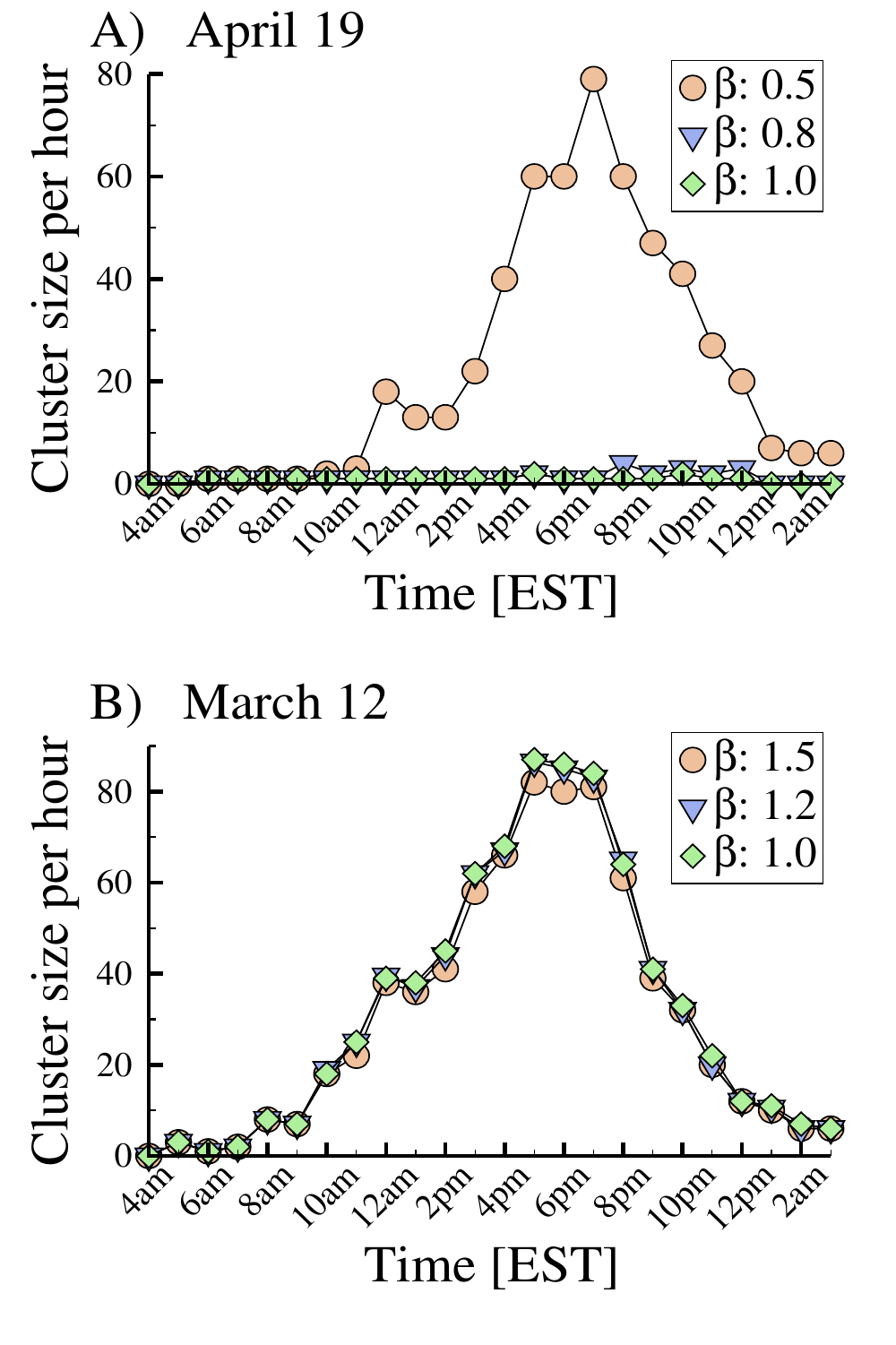}
\caption{Dependence of the hourly largest cluster with aS variation in $\beta$. A) April 19 and B) March 12.}
\label{fig:BetaVar}
\end{center}
\end{figure}

For April $19$ Table~\ref{tab:beta} shows that only by cutting the scheduled capacity by half, the day will start to have systemic problems according to the size of the largest cluster. In the case of March $12$ the scheduled airport capacity is increased by $50\%$ and the results indicates that this increment does not change the overall picture. Furthermore, Figure~\ref{fig:BetaVar} B shows that increasing airports' capacity will not ease off the propagation of delays. The reason for this is that the main cause of delay spreading, flight 
connections within the schedule, is independent of the airport capacity. Conversely, by reducing $\beta$ by at least $50\%$ can worsen the situation (Figure~\ref{fig:BetaVar} A). Such decrease on the airports' capacity can act as a trigger to new primary delays (different from the initial ones) that later on will spread in a cascading effect due to the flight connectivity. Although, a decrease on the scheduled capacity of $50\%$ for every airport in the network is not likely to occur in practice, a much realistic situation could be an airport or
group of airports operating undercapacity when severe weather conditions are met.

In any case, airport congestion could be the source for primary delays but it does not seem to be an important force behind their network-wide propagation.

\subsection{Stochastic variability of the results}
\label{sec:ModelErrors}

\begin{figure}[h]
\begin{center}
\includegraphics[angle=-90,width=8cm]{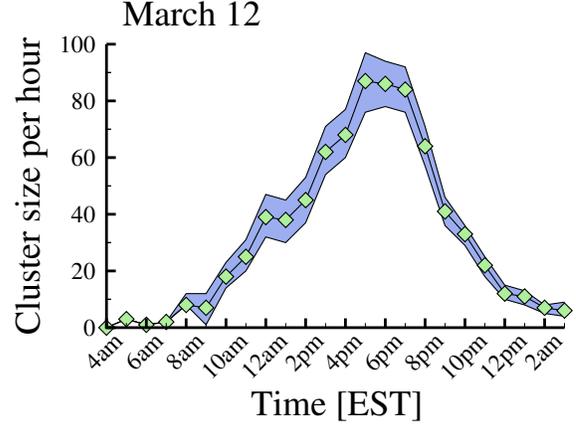}
\caption{Exploring the variability of the model results.}
\label{fig:variability}
\end{center} 
\end{figure}

Because of the stochasticity included in the model each realization has a slightly different outcome.
Figure~\ref{fig:variability} displays the variability between model realizations of the results for March $12$ considering a confidence interval of $95\%$. Simulations in this case were done using initial conditions "from the data"; this means that the stochasticity is caused only by flight connectivity. No matter which set of flight connections are randomly selected, March 12 will continue to display a
large cluster.  

In Figure~\ref{fig:variability} we can differentiate a growing phase that goes from $4$am to $5$pm and a declining phase from $5$pm onwards. As already said, merging is critical for the size evolution of the clusters. Because in the first hours of an unsatisfactory day there are several clusters, thus more possible combinations of merging events, the growing phase is characterized by a 
stronger variability than the declining phase. The latter, depicts a low variability and as Figure~\ref{fig:NumberClusters} A shows the number of clusters do not increase during this phase. All in all, this indicates that no atomization into smaller clusters is produce when the size diminishes. The cluster size dissolves continuously. 

\subsection{Further results on cluster and individual airport dynamics}
\label{sec:NumberClusters}

\begin{figure}[t!]
\begin{center}
\includegraphics[width=8cm]{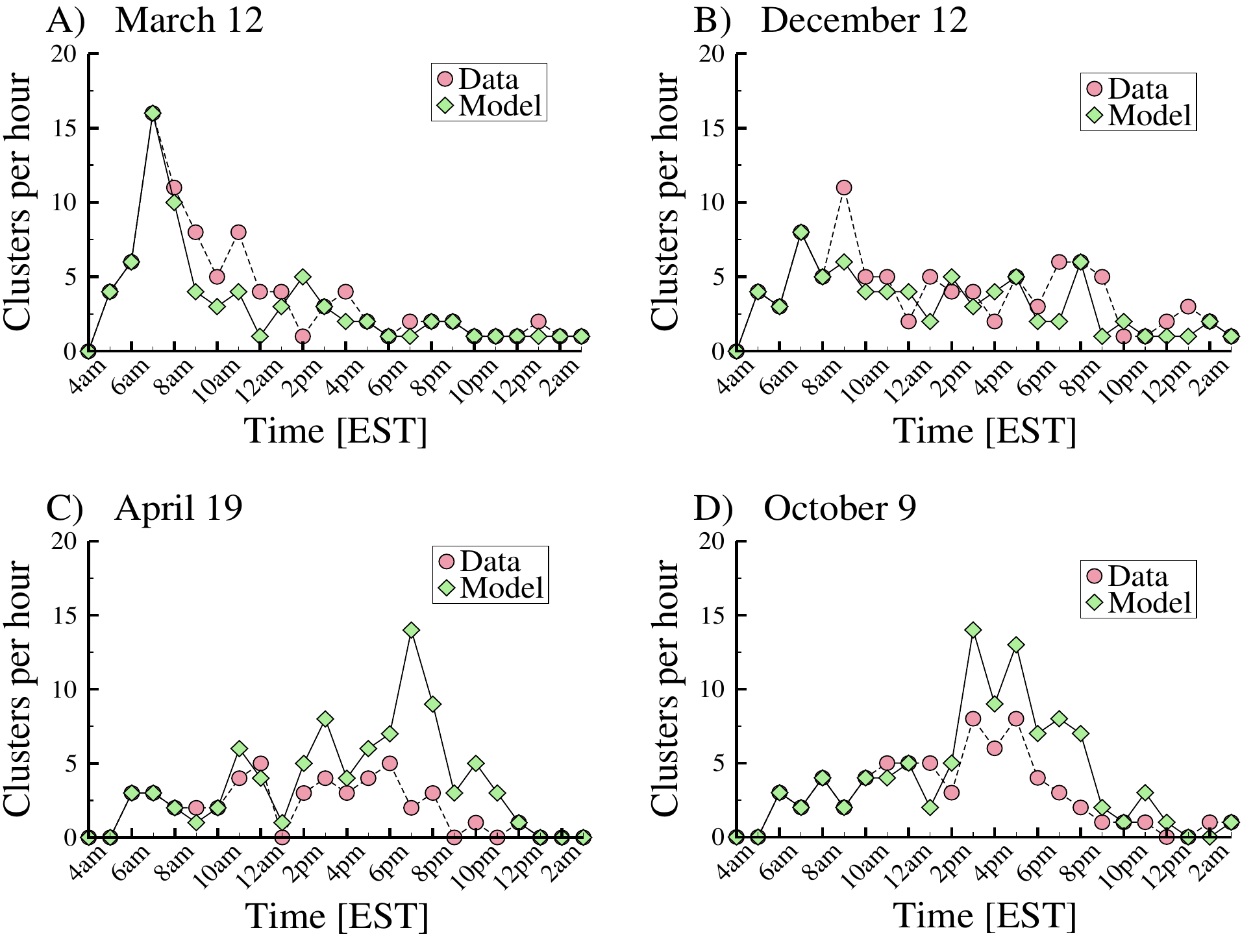}
\caption{Evolution of the number of clusters. Comparison between data and model results for: A) March 12, B) December 12, C) April 19 and D) October 9.}
\label{fig:NumberClusters}
\end{center}
\end{figure}

\begin{figure}
\begin{center}
\includegraphics[width=8cm]{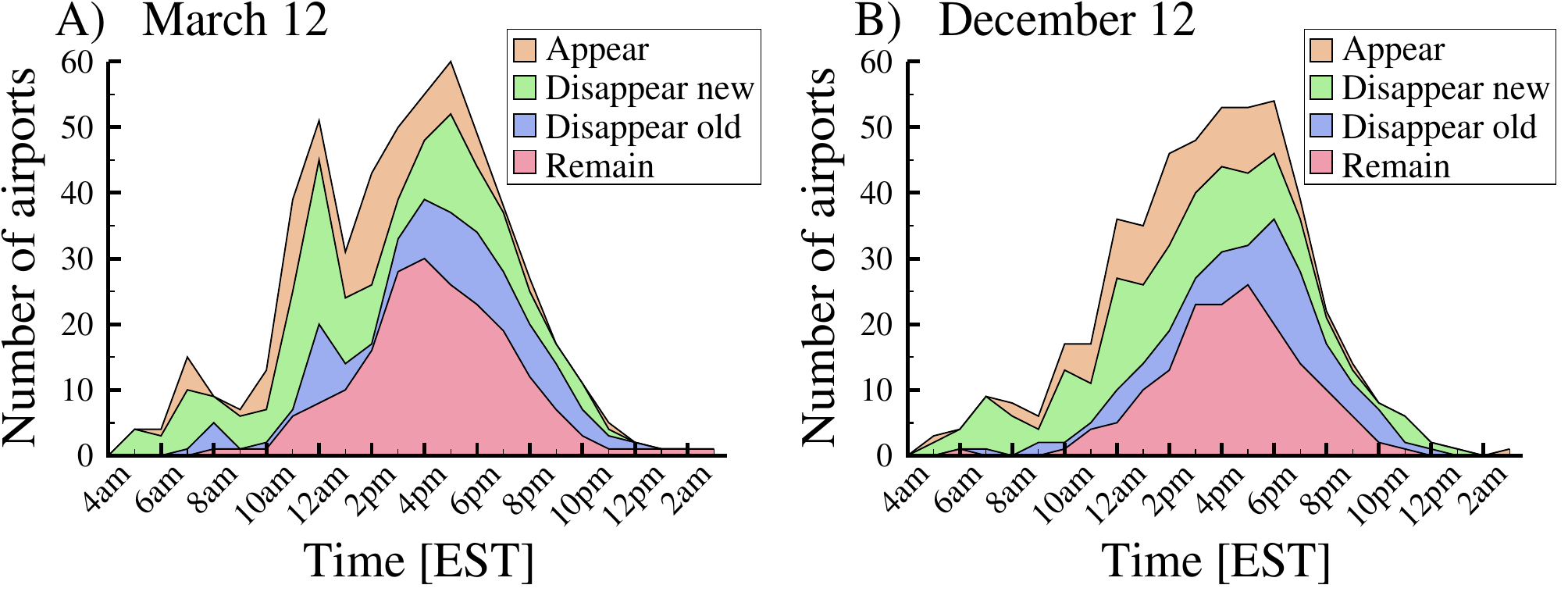}
\caption{Results obtained from the data. Number of airports that belongs to the largest cluster of the day for A) March 12 and B) December 12. Red color indicates the number of "old" airports (that 
their average departure delay per flight has been $>$ 29 minutes at least in the previous hour as well), while new airports that match this condition are shown in orange. The nodes which their 
average departure delay per flight will drop below 29 minutes in the next hour are shown in green (if they have been in problem for only one hour) and blue (if they have been in problem at least for 
two hours).}
\label{fig:GrowthDynamics}
\end{center}
\end{figure}

\begin{table}
\begin{center}
\begin{tabular}{c|c||c|c}
Airport code & days in  & Airport code & days with  \\
$\,$ & largest cluster & $\,$ & problems \\
\hline
\hline
ACV & $100$  & OTH & $167$\\
CEC & $80$  & CEC & $138$\\
SFO & $54$  & ACV & $136$\\
OTH & $52$  & LMT & $111$\\
MOD & $49$  & MOD & $90$\\
EWR & $45$ & CIC & $86$\\
CIC & $45$ & MFR & $70$\\
LMT & $44$ & BRW & $62$\\
MFR & $43$ & CRW & $60$\\
CRW & $41$ & MLB & $60$\\
\hline
\hline
\end{tabular}
\end{center}
\caption{Top 10 raking of airports in number of days belonging to the largest congested cluster or in number of days with problems.}
\label{tab:topdelay}
\end{table}

Besides the evolution of the size of the largest cluster per hour, dynamics can be characterized by the evolution of the number of clusters during the day (see Figure~\ref{fig:NumberClusters}). While in a satisfactory day (Figure~\ref{fig:NumberClusters} C and D) the number of clusters varies in each hour without a recognizable pattern, in an unsatisfactory day (Figure~\ref{fig:NumberClusters} A and B) the number of clusters increase in the first hours of the morning and from then on decay merging into fewer clusters, in most cases, in the afternoon (eastern time). This high-level interaction dynamics between clusters appears to be crucial in the evolution of an unsatisfactory day, where high-degree nodes play an important role to make this merging event come about. 

However events involving individual nodes occur and varies dramatically from time to time. Figure~\ref{fig:GrowthDynamics} displays how nodes that belong to the largest cluster of the day vary their condition rapidly. One hour they are above the $29$ minutes threshold and next they recover and vice versa. Most nodes switch from one state to the other very quickly, although some few nodes repeat their condition at least two time steps (red series).

In order to study the temporal persistence of airports in the largest congested cluster across the whole database, we display in Table~\ref{tab:topdelay} the list of the top 10 airports in days in the largest congested cluster and in days with problems. Although some airports appear in both lists, the order changes and both sets are not exactly equal. In these lists, there is a strong component of airports located in the West Coast. We think that this is due to the time difference between East and West Coasts. Flight operations initiate before in the East Coast and so the delays can propagate Westwards toward the end of the day. In the results in the main paper, we show that the largest congested clusters are not persistent, at least not in more than $ \sim 50\%$ of the airports between different days. The airports in Table~\ref{tab:topdelay} are those most persistent in the largest congested cluster. It is interesting to notice that only two major hubs, Newark (EWR) and San Francisco (SFO), are present in the top 10 list.

\end{document}